\documentclass[pra,showpacs,twocolumn,superscriptaddress]{revtex4}
\newcommand{\FigDirectory}{.}

\usepackage{amsmath}                          
\usepackage{amsfonts}                         
\usepackage{amssymb}                          
\usepackage{amscd}                            
\usepackage{amsthm}                           

\usepackage{dsfont}                           
\usepackage[usenames]{color}                  
\usepackage[OT2,T1]{fontenc}                  

\usepackage{float}                            
\usepackage{graphicx}                         
\usepackage{epstopdf}                         
\usepackage{subfigure}			              
\usepackage{comment}

\usepackage[colorlinks=true,
            urlcolor=blue,
            citecolor=blue,
            linkcolor=black,
            pdfstartview=FitH,
            pdfpagemode=None]{hyperref}

\usepackage{datetime}                         
\usepackage{longtable}
\usepackage{multirow}  
\usepackage{pbox}  
\usepackage{tabulary} 
\usepackage{algorithm}
\usepackage{algorithmic}
\usepackage{enumerate}

\renewcommand{\>}{\rangle}

\newcommand{\lsim}{\mathrel{\raise.3ex\hbox{$<$\kern-.75em\lower1ex\hbox{$\sim$}}}}
\newcommand{\gsim}{\mathrel{\raise.3ex\hbox{$>$\kern-.75em\lower1ex\hbox{$\sim$}}}}

\def\QECCnk[[#1,#2]]{[\![#1, #2]\!]}

\def\QECCnkq[[#1,#2,#3]]{[\![#1, #2]\!]_{#3}^{\vphantom{T}}}

\def\QECCnkd[[#1,#2,#3]]{[\![#1, #2, #3]\!]}

\def\QECCnkdq[[#1,#2,#3,#4]]{[\![#1, #2, #3]\!]_{#4}^{\vphantom{T}}}

\def\QECCnkgd[[#1,#2,#3,#4]]{[\![#1, #2, #3, #4]\!]}

\def\QECCnkgdq[[#1,#2,#3,#4,#5]]{[\![#1, #2, #3, #4]\!]_{#5}^{\vphantom{T}}}

\def\QECCnkdc[[#1,#2,#3,#4]]{[\![#1, #2, #3; #4]\!]}

\def\QECCnkdcq[[#1,#2,#3,#4,#5]]{[\![#1, #2, #3; #4]\!]_{#5}^{\vphantom{T}}}

\def\QECCnkgdcq[[#1,#2,#3,#4,#5,#6]]{%
  [\![#1, #2, #3, #4; #5]\!]_{#6}^{\vphantom{T}}}

\newcommand{\bigO}{{\cal O}}

\newcommand\CNOT{\ensuremath{\textit{CNOT\/}}}

\newcommand\SWAP{\ensuremath{\textit{SWAP\/}}}

\def\openone{\leavevmode\hbox{\small1\normalsize\kern-.33em1}}

\newcommand{\etal}{\textit{et~al.}}

\newcommand{\ie}{\textit{i.e.}}

\newcommand{\eg}{\textit{e.g.}}

\long\def\symbolfootnote[#1]#2{\begingroup%
\def\thefootnote{\fnsymbol{footnote}}\footnote[#1]{#2}\endgroup}

\input{Qcircuit.tex}

\theoremstyle{definition}  

\makeatletter
\def\grabtimezone #1#2#3#4#5#6#7#8#9{\grabtimezoneB}
\def\grabtimezoneB #1#2#3#4#5#6#7{\grabtimezoneC}
\def\grabtimezoneC #1#2'#3'{#1#2#3 UTC}
\def\timezone{\expandafter\grabtimezone\pdfcreationdate}
\makeatother

\begin{document}


%
\title{Quantum computing by color-code lattice surgery}

\author{Andrew J. \surname{Landahl}}
\email[]{alandahl@sandia.gov}
\author{Ciar\'{a}n \surname{Ryan-Anderson}}
\email[]{ciaranra@unm.edu}
\affiliation{Advanced Device Technologies,
             Sandia National Laboratories,
             Albuquerque, NM, 87185, USA}
\affiliation{Center for Quantum Information and Control,
             University of New Mexico,
             Albuquerque, NM, 87131, USA}
\affiliation{Department of Physics and Astronomy,
             University of New Mexico,
             Albuquerque, NM, 87131, USA}



\begin{abstract}

We demonstrate how to use lattice surgery to enact a universal set of
fault-tolerant quantum operations with color codes.  Along the way, we also
improve existing surface-code lattice-surgery methods.  Lattice-surgery
methods use fewer qubits and the same time or less than associated
defect-braiding methods.  Furthermore, per code distance, color-code lattice
surgery uses approximately half the qubits and the same time or less than
surface-code lattice surgery.  Color-code lattice surgery can also implement
the Hadamard and phase gates in a single transversal step---much faster than
surface-code lattice surgery can.  Against uncorrelated circuit-level
depolarizing noise, color-code lattice surgery uses fewer qubits to achieve
the same degree of fault-tolerant error suppression as surface-code lattice
surgery when the noise rate is low enough and the error suppression demand
is high enough.

\end{abstract}

%
\maketitle


%
\section{Introduction}

Planar topological quantum error-correcting codes have emerged as promising
substrates for fault-tolerant quantum computing because of their high
thresholds~\cite{Stephens:2014b}, compatibility with two-dimensional (2D)
local quantum processing~\cite{Dennis:2002a}, low quantum circuit overheads
\cite{Raussendorf:2007a}, efficient decoding algorithms~\cite{Dennis:2002a,
Duclos-Cianci:2009a, Stephens:2014a}, and the ability to smoothly
interpolate between desired effective error rates, which concatenated codes
cannot do~\cite{Cross:2009a}.  By Anderson's classification
theorem~\cite{Anderson:2011a}, the only alternatives for planar topological
stabilizer codes with nonlocal logical operators are surface
codes~\cite{Kitaev:1997a} and color codes~\cite{Bombin:2006b}.

In principle, fault-tolerant quantum computing with surface codes can be
achieved with transversal methods~\cite{Dennis:2002a}, defect-based
methods~\cite{Raussendorf:2006a, Raussendorf:2007a, Raussendorf:2007b}, or
lattice-surgery-based methods~\cite{Horsman:2012a}.  On 2D arrays of qubits
restricted to local quantum processing and local qubit movements,
transversal methods require an amount of information swapping that scales
with the system size.  Defect and lattice-surgery methods avoid this,
improving both their runtime and their accuracy
threshold~\cite{Spedalieri:2009a}.  Of these latter two, lattice surgery
uses substantially fewer qubits to achieve a desired error rate.  For
example, the fewest-qubit fault-tolerant distance-three $\CNOT$ method in a
topological code reported to date uses surface-code lattice surgery and only
requires 53 qubits~\cite{Horsman:2012a}.

Extending transversal surface-code methods to color codes is
straightforward.  Fowler has also extended defect-based surface-code methods to
defect-based color-code methods~\cite{Fowler:2008c}.  Notably absent are
extensions of surface-code lattice-surgery methods to color-code
lattice-surgery methods.  Developing such methods is especially important
because not only are lattice-surgery methods more qubit-efficient than
defect-based methods, but also color codes are significantly more
qubit-efficient than surface codes---for example, 4.8.8 color codes use
about half the qubits as the qubit-optimal medial surface
code~\cite{Bombin:2007d} to achieve the same code
distance~\cite{Landahl:2011a}.

Going beyond the application of a topological quantum
memory~\cite{Dennis:2002a}, color-codes offer additional advantages.  While
transversal two-qubit operations incur penalties for swapping information
around, one-qubit transversal operations do not; these advantages carry over
to lattice-surgery methods.  Two especially noteworthy methods are those for
the encoded, or ``logical,'' Hadamard gate ($H$) and those for the logical
phase gate ($S$) on planar color codes on the 4.8.8 lattice---both can be
implemented in a single parallelized transversal step~\cite{Bombin:2006b}.
For surface codes, neither of these gates have transversal implementations
on any lattice.  Current surface-code solutions for these gates include
elaborate multi-step code deformation procedures to implement the Hadamard
gate~\cite{Fowler:2012c, Horsman:2012a} and lengthy multi-gate teleportation
procedures from (previously distilled) magic states to implement the
phase gate~\cite{Raussendorf:2006a, Aliferis:2007b}.

The only downside to color codes versus surface codes is their lower
accuracy threshold, whose value has been estimated to be $0.143\%$ against
depolarizing circuit-level noise using a perfect-matching
decoder~\cite{Stephens:2014a}.  Surface codes have an accuracy threshold
whose value has been estimated to be in the range $0.502(1)\%$ to
$1.140(1)\%$~\cite{Stephens:2014b} in the same setting.  That said, surface
codes have enjoyed far greater study than color codes and we expect that
there are opportunities to close the gap.  We will show later that, even as
things stand now, at sufficiently low error rates and sufficiently low
desired error rates to be achieved by encoding, color codes still use fewer
qubits, despite their lower accuracy threshold.

Bolstered by the possibility of significant time and qubit reductions for
fault-tolerant operations, in this article we develop methods for universal
fault-tolerant quantum computation using color-code lattice surgery.  We
show that our methods use fewer qubits per logical operation than
surface-code lattice-surgery methods, including the smallest distance-three
$\CNOT$ in a topological code---our color-code lattice-surgery methods only
use 30 qubits when one allocates one syndrome qubit per face (or 22 if one
uses a single mobile syndrome qubit).  Along the way, we also improve the
surface-code lattice-surgery methods so that the distance-three $\CNOT$ now
only uses 39 qubits when one allocates one syndrome qubit per face (or 28 if
one uses a single mobile syndrome qubit).

In Sec.~\ref{sec:background}, we provide a brief background on triangular
4.8.8 color codes to help make our exposition better self-contained.  In
Sec.~\ref{sec:universal}, we describe fault-tolerant color-code
lattice-surgery methods for performing each element in a universal set of
operations.  In Sec.~\ref{sec:resource-analysis}, we calculate the circuit
width and depth overheads required by these methods and compare them to the
corresponding overheads required by surface-code lattice-surgery methods.
Sec.~\ref{sec:conclusion} concludes.

%
\section{Background}
\label{sec:background}

\begin{figure}[h!]
\center{
\subfigure[\ $d = 3$
lattice]{\includegraphics[width=0.3\columnwidth]{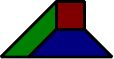}}\hspace{0.1em}
\subfigure[\ $d = 5$
lattice]{\includegraphics[width=0.3\columnwidth]{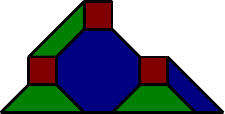}}\hspace{0.1em}
\subfigure[\ $d = 7$
lattice]{\includegraphics[width=0.3\columnwidth]{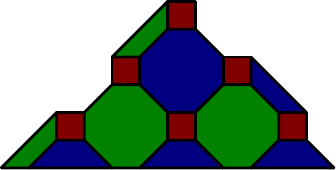}}
}
\caption{Triangular 4.8.8 color codes of distances 3, 5, and 7.  The number
of data qubits for distance $d$ is $(d^2-1)/2+d$.  The number of faces
(which is half the number of checks) is $(d^2+2d-3)/4$.
\small{\label{fig:color-codes} }}
\end{figure}

Our color-code lattice-surgery methods are valid for any color code, but for
concreteness we focus on lattice surgery of triangular color codes on the
4.8.8 lattice, namely the semiregular lattice that has a square and two
octagons surrounding each vertex.  These quantum stabilizer codes
\cite{Gottesman:1997a} exist for any odd code distance $d$ and can be
depicted graphically as in Fig.~\ref{fig:color-codes}.  Each vertex in this
figure corresponds to one (``data'') qubit in the code.  Each face in the
figure corresponds to two code checks, or stabilizer generators; one check
acts as Pauli $X$ on all qubits incident on the face and one check acts as
Pauli $Z$ on all qubits incident on the face.  The collection of qubits and
checks encode a single ``logical'' qubit.  Representatives of the logical
$X$ and $Z$ operators are strings of $X$ and $Z$ operators acting on the
qubits along the bottom side of the triangle.  By multiplying by a suitable
collection of check operators, two other equivalent representatives are
similar strings along either of the other two triangle sides.

Syndrome qubits are associated with the faces in the graph; how many
syndrome qubits are associated with each face is a nuanced function of the
syndrome extraction protocol one uses.  At a minimum, one can use a single
syndrome qubit over and over again, but it would have to be moved either
physically or by $\SWAP$ gates in such a way that it interacted with every
data qubit on the interior six times, every data qubit on the edge four
times, and every data qubit on a corner twice, because that is the number of
checks each of these types of data qubits are involved in.  A faster
syndrome extraction is possible by allocating one syndrome qubit per face so
that each syndrome qubit is used to measure both the $X$ and the $Z$ check
on each face.  By allocating two qubits per face, syndrome extraction can
run faster still, with the $X$ and $Z$ check measurements scheduled in an
interleaved fashion~\cite{Landahl:2011a}.

Adding more syndrome qubits can lead to better performance, such as a higher
accuracy threshold or less error propagation; we examine these tradeoffs in
greater detail in Sec.~\ref{sec:resource-analysis}.  One way to increase the
number of syndrome qubits is to allocate five syndrome qubits per each
octagonal face and two per each square face, extracting the syndrome into
two-qubit and verified four-qubit cat states~\cite{Fowler:2008c,
Stephens:2014a}.  By doubling this number of syndrome qubits, two cat states
per face can be prepared in parallel and used in the interleaved schedule
for $X$ and $Z$ check measurements.  Going even further, one can allocate
one syndrome qubit for every data qubit to enact even more robust
Shor-style~\cite{Shor:1996a} or Steane-style~\cite{Steane:1998a} syndrome
extraction.  This number of qubits can be doubled further to enact
Knill-style syndrome extraction with the same robustness but a faster
extraction circuit~\cite{Knill:2004a}.  We are not aware of any schemes that
use even more syndrome qubits to any advantage, so the number of syndrome
qubits can range anywhere from one to twice the number of data qubits.  In
this article, we will generally restrict attention to schemes which use
either one syndrome qubit per face or one syndrome qubit per check (two per
face), as we believe these offer the closest comparison to the most
widely-studied surface-code syndrome layout scheme, namely the one with one
syndrome qubit per check (one per face)~\cite{Dennis:2002a}.

Color codes are frequently considered in one of three broad classes of error
models~\cite{Landahl:2011a}.  In code-capacity models, data qubits are
subject to error but syndrome qubits are not.  In phenomenological models,
both data and syndrome qubits are subject to error.  In circuit-level
models, data qubits, syndrome qubits, and the individual quantum gates that
act upon them are subject to error.  This latter class is the most realistic
and is the one we focus on in this article.  However, because the available
operations at the circuit level are very hardware-dependent, we abstract
away the specifics of the hardware-level gate basis wherever possible.

Even when the physical circuit gate basis is known, it can be the case that
the error model on that gate basis is not well known.  In the absence of an
experimentally-informed circuit-level error model, a frequently used
surrogate is the independent identically distributed (iid) depolarizing
noise model, as it is kind of a ``worst case'' noise model for iid
stochastic errors.  In the iid depolarizing noise model, noise acts
independently and identically on the outputs of each quantum circuit
element, including the identity gate.  Depolarizing noise causes an error to
occur with probability $p$, and it selects the error equiprobably among the
possible non-identity Pauli operators on the outputs.  For single-qubit
measurement operations, it also flips the classical bit output with
probability $p$ (because a measurement error is a disagreement between the
recorded measurement outcome and the actual state).  While this noise model
is not without its flaws even for iid stochastic errors (see, for example,
Refs.~\cite{Magesan:2013a, Gutierrez:2013a, Stephens:2014b}), it is widely
used.

The syndrome extracted from a color code can be decoded in myriad ways.  For
the best performance, one could use the optimal decoder.  Although optimal
decoding of stabilizer codes is \#P-hard in general~\cite{Iyer:2013a}, it is
possible that an efficient optimal decoder (or one that approximates it
arbitrarily well) for color codes will be found.  For example, the
optimal-decoder-approximating PEPS decoder for surface-codes might be
extended to color codes~\cite{Bravyi:2014a}.  Alternatively, one could use a
slightly weaker integer-program-based decoder that identifies the most
likely error given the syndrome~\cite{Landahl:2011a}.  Weaker still but
faster yet, one could use a matching-based decoder, such as a minimum-weight
perfect matching decoder~\cite{Wang:2009b, Stephens:2014a}, a
renormalization-group matching decoder~\cite{Sarvepalli:2011a,
Duclos-Cianci:2011a, Duclos-Cianci:2014a}, a local greedy matching
decoder~\cite{Dennis:2003a, Bravyi:2013a, Wootton:2013a, Anwar:2014a}, or a
``global attractive-force'' local cellular automaton matching
decoder~\cite{Harrington:2004a, Herold:2014a}.  It is also possible to
exploit the local equivalence between a color code and a finite number of
copies of the surface code to arrive at a decoding solution from mulitple surface-code
decoders~\cite{Bombin:2012a, Delfosse:2014a}.  Developing new color-code
decoders is an active research front, where the trade space between decoding
complexity and decoding performance is being explored.

%
\section{Universal gate set}
\label{sec:universal}

In this section, we describe how to fault-tolerantly perform a universal set
of operations by lattice surgery on 4.8.8 triangular color codes.  We use
the notation from Ref.~\cite{Nielsen:2000a} to denote gates, states,
measurements, and quantum circuits.  The universal set we effect in
encoded form by lattice surgery is as follows:
\begin{align}
\left\{I, |0\>, |+\>, M_Z, M_X, S, H, T|+\>, \CNOT \right\}.
\end{align}

In the absence of hardware-informed circuit-level details, we imagine that
the same set of operations is available on the physical qubits as well, with
the $\CNOT$ gates restricted to nearest-neighbor data-ancilla qubit pairs.

With this gate basis, Pauli operators never need to be applied or even
synthesized from the other gates.  By the Gottesman-Knill
theorem~\cite{Gottesman:1999b}, Pauli operators can be propagated through
all stabilizer operations (Clifford gates plus Pauli preparations and
measurements) efficiently classically and used solely to reinterpret
measurement results.  Since this gate basis consists solely of stabilizer
operations and the $T|+\>$ preparation, and because Pauli operators never
need to be propagated through preparations, no Pauli operators are ever
needed.  Importantly, this means that if a decoding algorithm calls for
Pauli operators to be applied as a corrective action, the data need not be
touched by the Pauli operators and the classical ``Pauli frame'' can be
updated instead.  That said, to avoid polynomial-time classical computation,
it might be useful to implement the Pauli-frame updates from time to time.
For example, if errors are not corrected but only tracked, then the observed
syndrome bit rate will climb until it reaches a steady state close to 50\%,
at which point decoding may take longer than if the tracked Pauli
errors had been actually reversed.

In our fault-tolerant constructions, all but the $T|+\>$ preparation become
exponentially more tolerant to faults as the code distance increases.  To
increase the fidelity of $T|+\>$ preparations, any of a number of
magic-state distillation protocols can be used~\cite{Bravyi:2005a,
Meier:2012a, Bravyi:2012a, Landahl:2013a}.  These protocols use
high-fidelity operations from the rest of the set to ``distill'' multiple
$T|+\>$ preparations into fewer $T|+\>$ preparations of higher fidelity.

%
\subsection{The identity gate \texorpdfstring{$I$}{I}}

To fault-tolerantly implement the encoded identity gate on a triangular
color code, we simply perform fault-tolerant quantum error correction by
measuring the syndrome for $d$ rounds and run a classical decoding algorithm
on the data, such as one of the decoders described in
Refs.~\cite{Landahl:2011a, Duclos-Cianci:2014a, Stephens:2014a, Wang:2009b,
Sarvepalli:2011a}, to infer a corrective action.

%
\subsection{Preparation of \texorpdfstring{$|0\>$}{|0>} and
\texorpdfstring{$|+\>$}{|+>} states}

To fault-tolerantly prepare an encoded $|0\>$ state (the $+1$ eigenstate of
the encoded $Z$ operator), we first prepare each data qubit in a triangular
color code in the state $|0\>$ (the $+1$ eigenstates of the physical $Z$
operators).  We then perform fault-tolerant quantum error correction by
measuring the syndrome $d$ times and running it through a decoder.  The
process of measuring all of the code checks transforms the set of
single-qubit $Z$ checks into a set consisting of~($a$)~the $Z$ checks of the
color code and~($b$)~the encoded $Z$ operator for the color code.

The process for fault-tolerantly preparing an encoded $|+\>$ state (the $+1$
eigenstate of the encoded $X$ operator) is identical, except that the
individual qubits are initially prepared in $+1$ $X$ eigenstates instead of
$+1$ $Z$ eigenstates.

%
\subsection{Measurement \texorpdfstring{$M_Z$}{MZ} and
\texorpdfstring{$M_X$}{MX}}

To fault-tolerantly measure the encoded $Z$ operator, $M_Z$, on a logical
qubit, we measure each of the data qubits in the logical qubit in the $Z$
basis in a single round and perform classical error correction on the
result.  This measurement is ``destructive'' in that it takes the logical
qubit out of the code space.  A non-destructive measurement can be
implemented by augmenting this destructive measurement with an encoded
$\CNOT$ gate using Steane's ancilla-coupled measurement
method~\cite{Steane:1998a}.

Fault-tolerantly measuring the encoded $X$ operator, $M_X$, is similar: we
measure each of the data qubits in the logical qubit in the $X$ basis in a
single round and perform classical error correction on the result.  It is
also a destructive measurement, with a nondestructive version achievable
using Steane's method.

%
\subsection{Phase and Hadamard gates (\texorpdfstring{$S$}{S} and
\texorpdfstring{$H$}{H})}

Because the 2D color codes are \emph{strong} CSS codes (meaning that not
only do the checks factor into $X$-type and $Z$-type classes but also they
have identical support), the transversal Hadamard gate will swap the two
types of checks.  For triangular color codes (but not, \eg, for color codes
on compact surfaces \cite{Bombin:2007d}), the logical $X$ and $Z$ operators
can be made to be coincident so that the transversal Hadmard gate exchanges
these as well.  The net result is that the transversal Hadamard gate is a
fault-tolerant logical Hadamard gate for triangular color codes.

As shown by Bombin in Ref.~\cite{Bombin:2013a}, the $S$ gate is transversal
for 2D color codes as well, with a suitable choice of which physical qubits
to apply $S$ to and which to apply $S^\dagger$ to.  The 2D triangular color
codes on the 4.8.8 lattice have perhaps the simplest allocation choice: use
the transversal $S$ operator if the code distance is congruent to $1 \bmod
4$ and the transversal $S^\dagger$ operator if the code distance is
congruent to $3 \bmod 4$.

%
\subsection{The CNOT gate}

\begin{figure}[H]
%
%
\centerline{
\Qcircuit @C=1em @R=1em {
                         &                       & (-1)^b                 &            &                &     \\
 \lstick{\text{control}} & \qw                   & \multigate{2}{M_{ZZ}}  & \qw        & \gate{Z^{a+c}} & \qw \\
                         & (-1)^a                & \push{\rule{0em}{1.2em}} & (-1)^c     &                &     \\
 \lstick{\ket{0}}        & \multigate{2}{M_{XX}} & \ghost{M_{ZZ}}         & \gate{M_X} & \gate{Z^c}     & \qw \\
                         &                       & \push{\rule{0em}{1.2em}} &            &                &     \\
 \lstick{\text{target}}  & \ghost{M_{XX}}        & \qw                    & \qw        & \gate{X^b}     & \qw \\
} 
} 
%
\caption{\small{\label{fig:Bell-CNOT}Measurement-based $\CNOT$ circuit.}}
\end{figure}
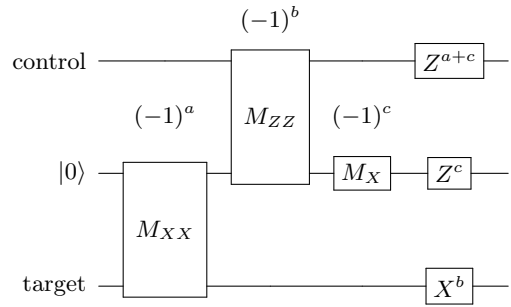

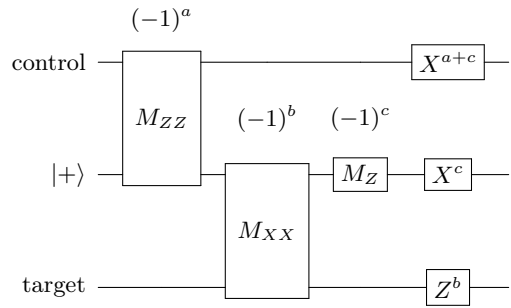
\begin{figure}[H]
%
%
\centerline{
\Qcircuit @C=1em @R=1em {
                         & (-1)^a                   &                          &            &                &     \\
 \lstick{\text{control}} & \multigate{2}{M_{ZZ}}    & \qw                      & \qw        & \gate{X^{a+c}} & \qw \\
                         & \push{\rule{0em}{1.2em}} & (-1)^b                   & (-1)^c     &                &     \\
 \lstick{\ket{+}}        & \ghost{M_{ZZ}}           &  \multigate{2}{M_{XX}}   & \gate{M_Z} & \gate{X^c}     & \qw \\
                         & \push{\rule{0em}{1.2em}} &                          &            &                &     \\
 \lstick{\text{target}}  & \qw                      &  \ghost{M_{XX}}          & \qw        & \gate{Z^b}     & \qw \\
} 
} 
%
\caption{\small{\label{fig:Horsman-CNOT}Alternative measurement-based $\CNOT$ circuit.}}
\end{figure}

To fault-tolerantly implement the encoded $\CNOT$ gate, we use a sequence of
lattice surgery operations.  These operations are intended to mimic either
the circuit in Fig.~\ref{fig:Bell-CNOT} or the circuit in
Fig.~\ref{fig:Horsman-CNOT}, both of which are equivalent to a $\CNOT$ gate;
these circuits were leveraged heavily in Ref.~\cite{Aliferis:2008b} to
combat biased noise.

The Pauli corrections in these circuits can be omitted in our approach 
because of our choice of gate basis; we simply use them to re-interpret
future measurement results as needed.  The only operations depicted in these
circuits that we have not provided methods for yet are the $M_{XX}$ and
$M_{ZZ}$ measurements; with them, we can construct the encoded $\CNOT$
operation.

\begin{figure}[H]
\begin{center}
  \includegraphics[width=0.75\columnwidth]{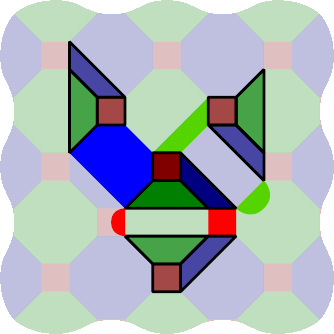}
\caption{\small{\label{fig:MXX-MZZ-d3}To measure $M_{XX}$ ($M_{ZZ}$) between the
central logical qubit and a logical qubit adjacent to one of its sides, measure
only the $X$ ($Z$) checks on the lighter-colored faces on the interface and
the $X$ and $Z$ checks on the full octagons shared across the interface.  (The
figure compresses three separate scenarios into one.)  The outcome is the
product of the lighter-colored check outcomes.  (color online.)}}
\end{center}
\end{figure}

\begin{figure}[H]
\begin{center}
  \includegraphics[width=0.75\columnwidth]{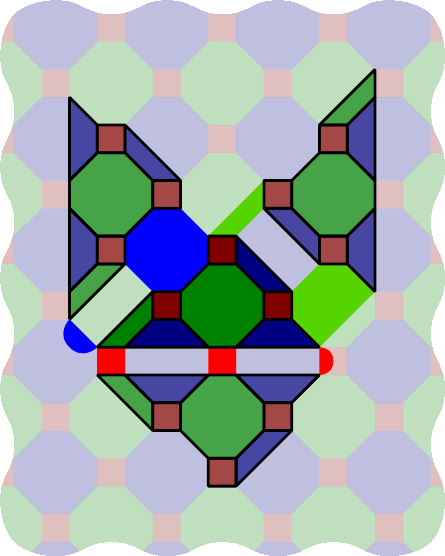}
\caption{\small{\label{fig:MXX-MZZ-d5}The same scenario as
Fig.~\ref{fig:MXX-MZZ-d3}, except with distance-five codes. (color online)}}
\end{center}
\end{figure}

To measure $XX$ or $ZZ$ between two triangular color codes, we measure
checks that connect the adjacent logical qubits in an ``osculating'' manner.
Figures~\ref{fig:MXX-MZZ-d3} and \ref{fig:MXX-MZZ-d5} depict how this can be
done for every side of a 4.8.8 triangular color code for code distances 3
and 5; the pattern generalizes in a straightforward way. 

\begin{figure}[h]
\begin{center}
  \includegraphics[width=0.75\columnwidth]{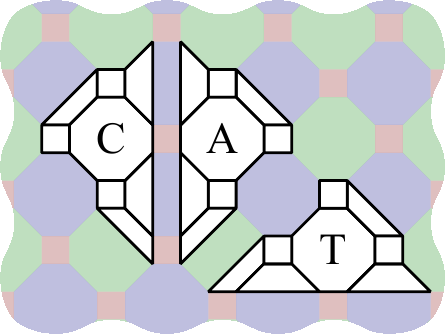}
\caption{\small{\label{fig:CAT-layout-d5}Regions outlined and filled with
white indicate where the control (C), ancilla (A), and target (T) qubits are
located for a distance-five example.  (color online)}}
\end{center}
\end{figure}

Using these methods for $M_{XX}$ and $M_{ZZ}$ measurements, we describe
step-by-step how to implement a fault-tolerant $\CNOT$ gate by lattice
surgery using a simulation of the circuit in Fig.~\ref{fig:Bell-CNOT}; the
simulation of the circuit in Fig.~\ref{fig:Horsman-CNOT} is similar.  While
our construction works for arbitrary code distances, we depict an example of
each step for $d=5$, with the layout of control, ancilla, and target regions
as depicted in Fig.~\ref{fig:CAT-layout-d5}; other choices of orientation
are possible.

\begin{figure}[H]
\begin{center}
  \includegraphics[width=0.75\columnwidth]{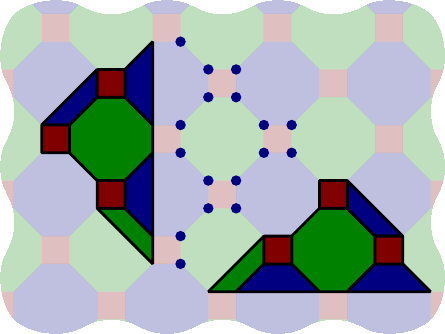}
\caption{\small{\label{fig:CAT-PZhat}(Step 1.) The qubits in the ancilla region (A) are
prepared in $Z = +1$ eigenstates.  (color online)}}
\end{center}
\end{figure}

\begin{figure}[H]
\begin{center}
  \includegraphics[width=0.75\columnwidth]{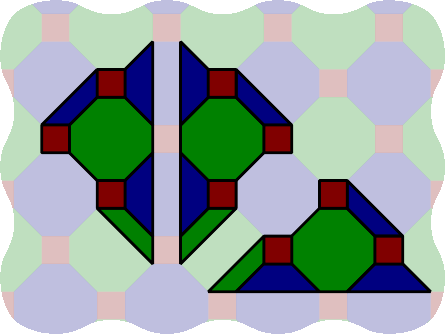}
\caption{\small{\label{fig:CAT-P0}(Step 2.) The checks in the ancilla region (A) are
measured for $d$ rounds and errors are corrected fault-tolerantly.  (color online)}}
\end{center}
\end{figure}

\begin{figure}[H]
\begin{center}
  \includegraphics[width=0.75\columnwidth]{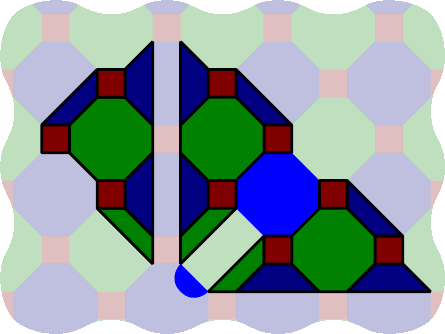}
\caption{\small{\label{fig:CAT-MXX}(Step 3.) The checks that fuse the target and
ancilla logical qubits in an $M_{XX}$ measurement are measured for $d$
rounds and errors are corrected fault-tolerantly. (color online)}}
\end{center}
\end{figure}

\begin{figure}[H]
\begin{center}
  \includegraphics[width=0.75\columnwidth]{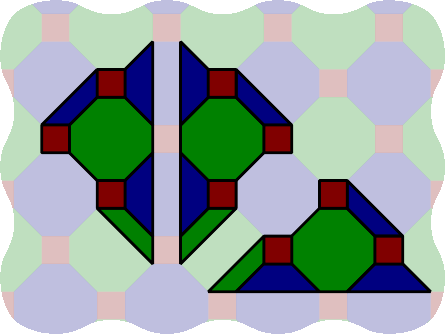}
\caption{\small{\label{fig:CAT-MXX-split}(Step 4.) The $M_{XX}$-fusing checks stop
being measured.  Instead, the target and ancilla logical qubits checks are
measured for $d$ rounds and errors are corrected fault-tolerantly.  (color
online)}}
\end{center}
\end{figure}

\begin{figure}[H]
\begin{center}
  \includegraphics[width=0.75\columnwidth]{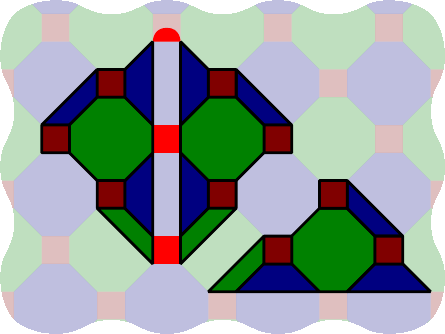}
\caption{\small{\label{fig:CAT-MZZ}(Step 5.) The checks that fuse the control and
ancilla logical qubits in an $M_{ZZ}$ measurement are measured for $d$
rounds and errors are corrected fault-tolerantly. (color online)}}
\end{center}
\end{figure}

\begin{figure}[H]
\begin{center}
  \includegraphics[width=0.75\columnwidth]{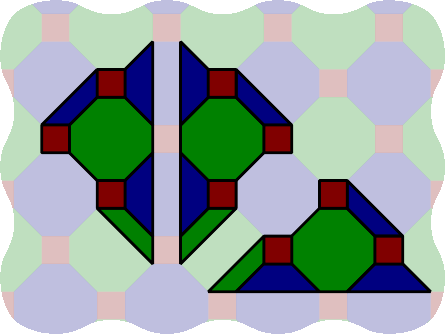}
\caption{\small{\label{fig:CAT-MZZ-split}(Step 6.) The $M_{ZZ}$-fusing checks stop
being measured.  Instead, the control and ancilla logical qubits checks are
measured for $d$ rounds and errors are corrected fault-tolerantly.  (color
online)}}
\end{center}
\end{figure}

\begin{figure}[H]
\begin{center}
  \includegraphics[width=0.75\columnwidth]{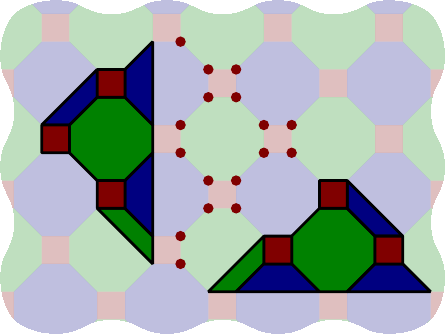}
\caption{\small{\label{fig:CAT-MX}(Step 7.) The qubits in the ancilla region are
measured in the $X$ basis, implementing a destructive $M_X$ measurement.
The result is error-corrected classically.  The control and target logical
qubit checks are measured for $d$ rounds and errors are corrected
fault-tolerantly. (color online)}}
\end{center}
\end{figure}

\begin{enumerate}
\item Prepare the data qubits in the ancilla region in $|0\>$ states ($Z =
+1$ eigenstates), as depicted in Fig.~\ref{fig:CAT-PZhat}.
\item Measure the checks in the ancilla region for $d$ rounds and correct
errors fault-tolerantly, as depicted in Fig.~\ref{fig:CAT-P0}.
\item Measure the checks that fuse the target and ancilla logical qubits in
an $M_{XX}$ measurement for $d$ rounds and correct errors fault-tolerantly,
as depicted in Fig.~\ref{fig:CAT-MXX}.
\item Stop measuring the $M_{XX}$-fusing checks and measure the checks for
the target and ancilla logical qubits separately, splitting them apart
again, for $d$ rounds and correct errors fault-tolerantly, as depicted in
Fig.~\ref{fig:CAT-MXX-split}.
\item Measure the checks that fuse the control and ancilla logical qubits in
an $M_{ZZ}$ measurement for $d$ rounds and correct errors fault-tolerantly,
as depicted in Fig.~\ref{fig:CAT-MZZ}.
\item Stop measuring the $M_{ZZ}$-fusing checks and measure the checks for
the control and ancilla logical qubits separately, splitting them apart
again, for $d$ rounds and correct errors fault-tolerantly, as depicted in
Fig.~\ref{fig:CAT-MZZ-split}.
\item Measure the data qubits in the ancilla region in the $X$ basis
destructively and perform classical error correction on the result, as
depicted in Fig.~\ref{fig:CAT-MX}.
\end{enumerate}

%
%

%
%

%
%

%
%

%
%

%
%

%
%

As described, this method takes one round of data-qubit preparation, $5d$
rounds of syndrome extraction, and one round of data-qubit measurement.
However, this time can be sped up considerably.

As a starter, a preparation operation on a logical qubit and a fusing
operation between that logical qubit and another logical qubit can be
combined into a single step---instead of thinking of the operations as
``prepare-then-fuse,'' one can think of them as a single ``grow one of the
logical qubits'' operation.  Step 2 can therefore be eliminated and, without
loss of generality, we can omit step 1 and use the state it prepares as the
initial state of the method.  This reduces the number of rounds of
parallelized measurements to $4d+1$.

Next, a splitting operation between two logical qubits that ``heals'' the
interface between them can happen simultaneously with a fusing operation
acting on a different side of one of the logical qubits and a side of a
third logical qubit.  Running these operations simultaneously does not
hamper the fault-tolerance of the method---the code distances do not drop by
this kind of parallelization.  This observation allows us to eliminate step
4, reducing the number of rounds of parallelized measurements to $3d+1$.  It
also means that the target logical qubit is free to use one of its other
sides after just $d$ rounds of measurements.

Finally, a splitting operation between two logical qubits can happen
simultaneously with a destructive measurement operation that follows on one of them;
again, the operations do not interfere with one another.  Because the
destructive measurement operation only takes one round of parallelized
measurements, the time savings is not very great---the number of rounds is
reduced to $3d$ with this observation.

%
\subsection{Preparation of \texorpdfstring{$T|+\>$}{T|+>} states}

\begin{figure}[H]
\center{
  \subfigure[\
Step~1]{\includegraphics[width=0.45\columnwidth]{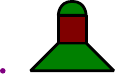}}\hspace{1em}
  \subfigure[\
Step~2]{\includegraphics[width=0.45\columnwidth]{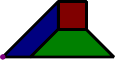}}
}
\caption{\small{\label{fig:d3-injection}Injection of $T|+\>$ qubit state
(purple dot) into  $d=3$ triangular 4.8.8 color code (image on right).  In steps 1
and 2, the indicated code checks are measured three times each.  (color online.)}}
\end{figure}

\begin{figure}[H]
\center{
  \subfigure[\
Step~1]{\includegraphics[width=0.45\columnwidth]{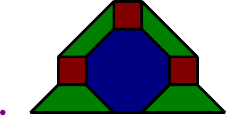}}\hspace{1em}
  \subfigure[\
Step~2]{\includegraphics[width=0.45\columnwidth]{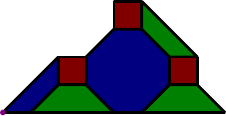}}
}
\caption{\small{\label{fig:d5-injection}Same as Fig.~\ref{fig:d3-injection},
but for a $d=5$ triangular 4.8.8 color code.  (color online.)}}
\end{figure}

\begin{figure}[H]
\center{
  \subfigure[\
Step~1]{\includegraphics[width=0.45\columnwidth]{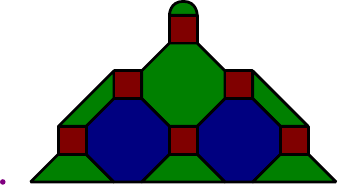}}\hspace{1em}
  \subfigure[\
Step~2]{\includegraphics[width=0.45\columnwidth]{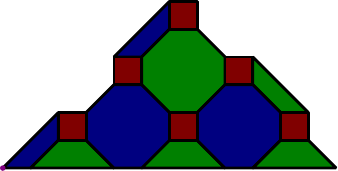}}
}
\caption{\small{\label{fig:d7-injection}Same as Fig.~\ref{fig:d3-injection},
but for a $d=7$ triangular 4.8.8 color code.  (color online.)}}
\end{figure}

To fault-tolerantly prepare an encoded $T|+\>$ state, we use the process of
code injection.  Figures~\ref{fig:d3-injection}--\ref{fig:d7-injection}
depict the injection process for distances $d=3$, $5$, and $7$.  The
coloring in these figures is chosen so that the blue side of the final
triangular code is always on the left for ease of discussion.  The top two
rows of qubits in these figures represent two isolated Bell pairs for $d=3$ and $d=7$,
even though they look like they are connected to the rest of the surface via
a square and a digon.

In the first step, we prepare a single qubit in the state $T|+\>$ and we
prepare an adjacent region in an auxillary state that consists of a distance
$d-1$ color-code stabilizer state, along with two additional Bell pairs if
$d \equiv 3 \bmod 4$.  For $d > 3$, we prepare the two Bell pairs to
$\bigO(p^2)$ error by post-selection, with a mean waiting time of
$(1-p)^{-4} \cong 1 + 4p$ rounds of measurement.  In parallel, we measure
the rest of the checks three times and use a classical decoding algorithm to
suppress errors in the distance $d-1$ code state to $\bigO(p^2)$.  We handle
the case of $d=3$ separately; the auxillary state is just three Bell states
in this case, so we prepare it by post-selection to $\bigO(p^2)$ error with
a mean waiting time of $(1-p)^{-6} \cong 1 + 6p$ rounds of measurement.

To inject the state, in the second step we measure the new blue $X$ and $Z$
checks along the left side, accepting whatever syndrome values we obtain as
being ``correct.'' This causes the Pauli $X$ and $Z$ operators on the single
qubit being injected to extend to distance-$d$ logical Pauli $X$ and $Z$
operators along that edge of the triangle.  In parallel, we cease measuring
the green checks along the left side, including the digon operator if one is
present.  However, in parallel we \emph{do} measure all of the other checks
for the code.

For $d > 3$, the checks that persist are capable of detecting up to two
errors on any pair of data qubits, excluding the state to be injected.  Any
single or two-qubit error on the interior data qubits will be detected
because the code distance is sufficiently high.  Any single-qubit error on
data qubits along the left boundary will be incident on a red check or the
bottom-left green check, so it will be detected as well.  If a two-qubit
error afflicts two data qubits on different red checks on the left side or a
red check and the bottom-left green check, they will also be detected.  If a
two-qubit error afflicts two data qubits on a single red check, at least one
other persistent check will detect it, by inspection.  Since the persistent
checks can detect up to two errors, one can use a classical decoding
algorithm on three rounds of extracted syndrome to correct any single error,
suppressing errors to $\bigO(p^2)$.  The case of $d=3$ can be handled as a
special case with, \eg, postselection on the entire injection process.

The total number of rounds of syndrome extraction in the state-injection
process is six: three to prepare the ancillary state and three to decode the
full distance-$d$ code.  The error in the process is $\bigO(p)$, where the
multiplicative constant is solely a function of the circuit elements in the
check measurement circuit that act on the state to be injected.
Importantly, this constant does not grow with the distance of the code.  To
reduce this error further once it is encoded, an encoded magic-state
distillation protocol may be used.

%
\section{Resource analysis}
\label{sec:resource-analysis}

%
\subsection{Overhead per code distance}
\label{sec:overhead-per-code-distance}

Table~\ref{tab:color-code-resources-2qubit} summarizes the space and time
resource overheads used by our color-code lattice-surgery methods for the
scenario in which one syndrome qubit is allocated per check (two per face).
\begin{center}
\begin{table}[h]
\centering
\begin{tabular}{c||c|c|c|c|c|c|c|c|c}
\multicolumn{10}{c}{Color-code lattice surgery (1 syndrome qubit/check)} \\[1ex] \hline \hline
Gate   & $T|+\>$ & $I$ & $|0\>$ & $|+\>$ & $M_Z$ & $M_X$ & $H$ & $S$ & $\CNOT$ \\[1ex] \hline
Depth  & $6$ & \multicolumn{3}{c|}{$d$} & \multicolumn{2}{c|}{$1$} & \multicolumn{2}{c|}{$0$}
       & $3d$ \\[1ex]  \hline
Qubits & \multicolumn{8}{c|}{$d^2 + 2d - 2$}
       & $3d^2 + 6d - 6$ \\[1ex] \hline
Error  & $\bigO(p)$ & \multicolumn{8}{c}{$\bigO(p^{(d+1)/2})$} \\[1ex] \hline \hline
\end{tabular}
\caption{Resources used by fault-tolerant 4.8.8 triangular color-code
lattice surgery on distance-$d$ codes when two syndrome bits per face are
allocated.  Depth is measured in number of measurement rounds.
Qubit counts include both data and syndrome qubits.  Error is reported in
big-$\bigO$ notation because syndrome-extraction-circuit implementation details
can change the constants.}
\label{tab:color-code-resources-2qubit}
\end{table}
\end{center}

\begin{figure}[h]
\begin{center}
  \includegraphics[width=0.75\columnwidth]{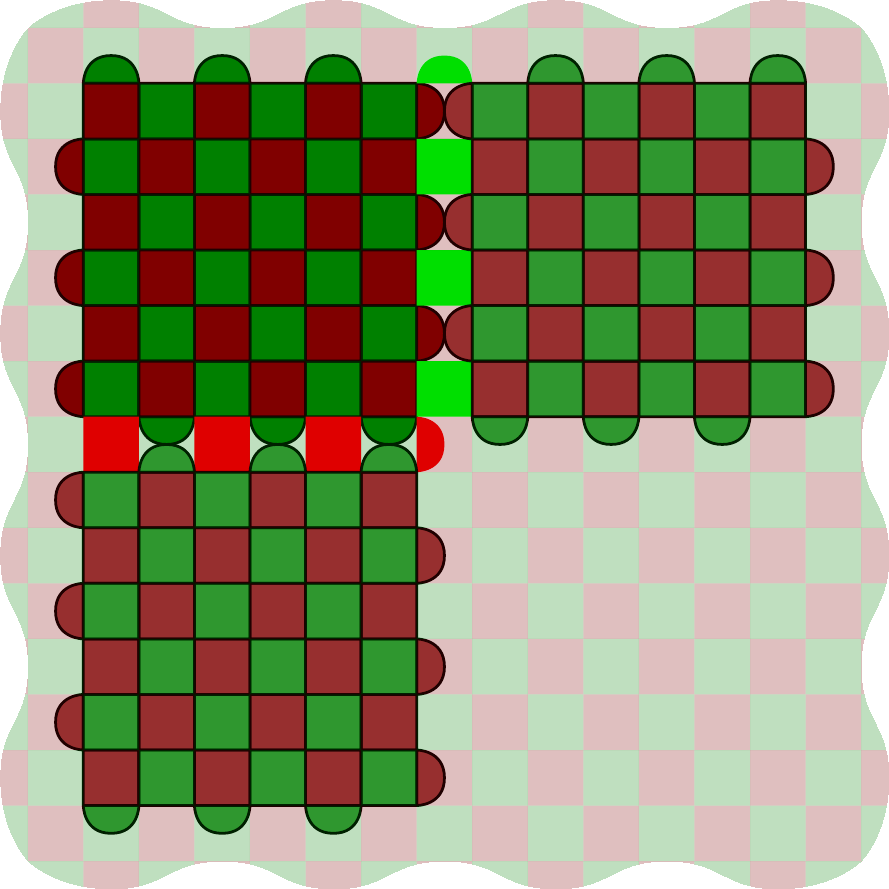}
\caption{\small{\label{fig:surface-CNOT}Layout for the $\CNOT$ gate on
surface codes as in Ref.~\cite{Horsman:2012a}, except with the intermediate
row of data qubits in the osculant regions removed.  The same layout is used
for the Hadamard gate, which grows and shrinks around the corner to
change the orientation of its boundary coloring. (color online)}}
\end{center}
\end{figure}

\begin{figure}[h]
\begin{center}
  \includegraphics[width=0.75\columnwidth]{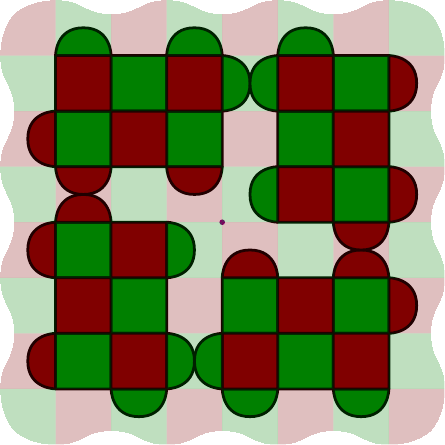}
\caption{\small{\label{fig:surf-code-chiral-injection}Injection procedure
for surface codes similar to the one in Fig.~\ref{fig:d7-injection}.  In
steps 1 and 2, the indicated code checks are measured three times each.  (color online)}}
\end{center}
\end{figure}

While surface-code lattice-surgery was first explored in by Dennis \etal\ in
the context of state injection~\cite{Dennis:2002a}, the first exploration of
a universal set of logical gates on surface codes using lattice-surgery
methods was performed by Horsman {\textit{et al.}}~\cite{Horsman:2012a}.
Inspired by our color-code lattice surgery methods, we improved the methods
presented in Ref.~\cite{Horsman:2012a} so that they now use fewer qubits for
the $\CNOT$, $H$, and $S$ gates, using the layout depicted in
Fig.~\ref{fig:surface-CNOT}.  We also developed a new six-step surface-code
state-injection method similar to our color-code state-injection method; the
surface-code layout is depicted in
Fig.~\ref{fig:surf-code-chiral-injection}.
Table~\ref{tab:surface-code-resources} lists the resources used by these
improved surface-surgery methods on the ``rotated'' or ``medial'' surface
code, with an allocation of one syndrome per check (one per face).

\begin{center}
\begin{table}[h]
\centering
\begin{tabular}{c||c|c|c|c|c|c|c|c|c}
\multicolumn{10}{c}{Surface-code lattice surgery (1 syndrome qubit/check)} \\[1ex] \hline \hline
Gate   & $T|+\>$ & $I$ & $|0\>$ & $|+\>$ & $M_Z$ & $M_X$ & $H$ & $S$ & $\CNOT$ \\[1ex] \hline
Depth  & $6$ & \multicolumn{3}{c|}{$d$} & \multicolumn{2}{c|}{$1$}
       & {$6d$}
       & {$12d$} & {$3d$} \\[1ex]  \hline
Qubits & \multicolumn{6}{c|}{$2d^2 -2d + 1$}
       & \multicolumn{3}{c}{$6d^2 - 6d+3$} \\[1ex] \hline
Error  & $\bigO(p)$ & \multicolumn{8}{c}{$\bigO(p^{(d+1)/2})$} \\[1ex] \hline \hline
\end{tabular}
\caption{Resources used by fault-tolerant medial surface-code lattice
surgery on distance-$d$ codes when one syndrome bit per face is allocated.
Depth is measured in number of measurement rounds.  Qubit counts include both
data and syndrome qubits.  Error is reported in big-$\bigO$ notation because
syndrome-extraction-circuit implementation details can change the constants.
The logical $S$ gate is implemented by catalytic teleportation from the $HS|+\>$
state, which requires two logical $\CNOT$ gates and a logical Hadamard gate
\cite{Aliferis:2007b}.  The logical $H$ gate is performed by lattice surgery as in
Ref.~\cite{Horsman:2012a}, but qubits are shifted $d$ sites horizontally and
$d$ sites vertically in the method to ensure that the size of the logical
operators do not drop below $d$, making the operation fault-tolerant.}
\label{tab:surface-code-resources}
\end{table}
\end{center}

From these tables, we see that color codes use approximately half as many
qubits as surface codes to achieve the same order of error suppression.
Color-code lattice surgery also performs encoded gates in essentially the
same time or faster than they are performed via surface-code lattice
surgery.  Even when both models are optimized for qubits by exploiting a
single roving syndrome qubit, the color-code $\CNOT$ uses $(3d^2 + 6d -
1)/2$ qubits whereas the surface-code $\CNOT$ uses $3d^2 + 1$
qubits---again about half as many.

%
\subsection{Overhead per desired level of error suppression}
\label{sec:overhead-per-error-suppression}

Because the accuracy threshold against circuit-level depolarizing noise is
smaller for color codes than for surface codes, a color code will need a
larger code distance than a surface would need to achieve the same
level of error suppression (\ie, to achieve the same logical failure
probability $p_{\text{fail}}$).  This erodes the factor-of-two qubit
savings that color codes provide at the same code distance, and could
possibly eliminate the savings entirely.

To compute the qubit overhead $\Omega$ to achieve a given $p_{\text{fail}}$
for a logical operation, one inverts the relationship $p_{\text{fail}}(d)$
and plugs the solution $d(p_{\text{fail}})$ into the appropriate expression
for the number of qubits per operation, \eg, from the ``Qubits'' entry in
Table~\ref{tab:color-code-resources-2qubit} or
Table~\ref{tab:surface-code-resources}.  The analytic expression best-suited
for $p_{\text{fail}}(d)$ depends on the relative magnitudes of $d$ and the
depolarizing probability $p$~\cite{Dennis:2002a, Wang:2003a,
Raussendorf:2007b, Fowler:2013a, Watson:2013a, Bravyi:2013a}; for example,
Watson and Barrett have shown that the scaling of $p_{\text{fail}}$ with $d$
is qualitatively different in the regime $d < 1/4p$ and $d > 1/4p$ for
code-capacity and phenomenological error models~\cite{Watson:2013a}.  Since
overhead comparisons are most relevant for non-asymptotic $d$ and for $p$
below the relevant pseudothreshold (\ie, the $p$ at a \emph{fixed} code
distance below which $p_{\text{fail}} < p$), and because we are most
interested in the scaling for circuit-level error models, we use the
expression for fixed $d$ and low $p$ for these models that Fowler found fit
well to surface-codes in Ref.~\cite{Fowler:2013a}, namely
\begin{align}
\label{eq:combinatoric-bound}
p_{\text{fail}} = A(d)\left(\frac{p}{p_{\text{th}}}\right)^{d/2}.
\end{align}

It is an interesting question as to whether color codes can exhibit the same
scaling at this in the low-$p$ regime.  Stephens has noted that his
color-code matching decoder in Ref.~\cite{Stephens:2014a} does not attain
the full algebraic code distance, suggesting that the exponent in
Eq.~(\ref{eq:combinatoric-bound}) using his decoder will be $\alpha d$,
where $\alpha < 1/2$.  In contrast, the integer-program (IP) decoder in
Ref.~\cite{Landahl:2011a} should attain the full code distance at the cost
of running more slowly.  If only one syndrome qubit per face or one per
check is used with the IP decoder, though, errors may spread badly, cutting
in to the effective code distance.  Using Shor-, Steane-, or Knill-style
syndrome extraction should eliminate this problem at the cost of many extra
syndrome qubits.  It may suffice to use the verificed four-cat and two-cat
states per octagonal and square faces respectively as used in
Refs.~\cite{Fowler:2008c, Stephens:2014a} with the IP decoder to achieve
this scaling, but currently that is an open question.  Although the IP
decoder appears to be inefficient at high error rates, at low error rates it
can be expected to run quickly.  Moreover, the recent linear-time PEPS
decoder for surface codes \cite{Bravyi:2014a} gives hope that a truly
efficient color-code decoder that achieves the scaling of
Eq.~(\ref{eq:combinatoric-bound}) will be found.  For the purposes of
comparision, and with this optimism in mind, we will assume that the scaling
law in Eq.~(\ref{eq:combinatoric-bound}) holds for both surface and color
codes.  However, we urge caution in reading too much into the results
derived from this assumption.

Using Eq.~(\ref{eq:combinatoric-bound}), the color-code distance $d_c$ that
gives the same error-suppression power as a surface code with distance $d_s$
is 
\begin{align}
\label{eq:dc-to-ds}
d_c &=
  d_s\left(\frac{\log p/p_{\text{th}}^{(s)}}
     {\log p/p_{\text{th}}^{(c)}}\right)
  + 2\left(\frac{\log A_s(d)/A_c(d)}
     {\log p/p_{\text{th}}^{(c)}}\right).
\end{align}
Fowler's numerical simulations suggest that $A_s(d)$ is approximately a
constant function of $d$ for $d$ up to 10~\cite{Fowler:2013a}; there is no
reason to expect that $A_c(d)$ is not also a comparably-sized constant
function of $d$ in the same range, or indeed that $A_s(d)$ and $A_c(d)$
should scale substantially differently for any $d$.  The numerator in the
second term of Eq.~(\ref{eq:combinatoric-bound}) should therefore be quite
small because of the logarithm.  Moreover, the denominator gets larger as
$p$ is reduced below the color-code (pseudo)threshold, making the overall
term even smaller.  For these reasons, we will neglect the second term in
Eq.~(\ref{eq:combinatoric-bound}) in our subsequent analysis.

\begin{figure}[h]
\begin{center}
  \includegraphics[width=1.0\columnwidth]{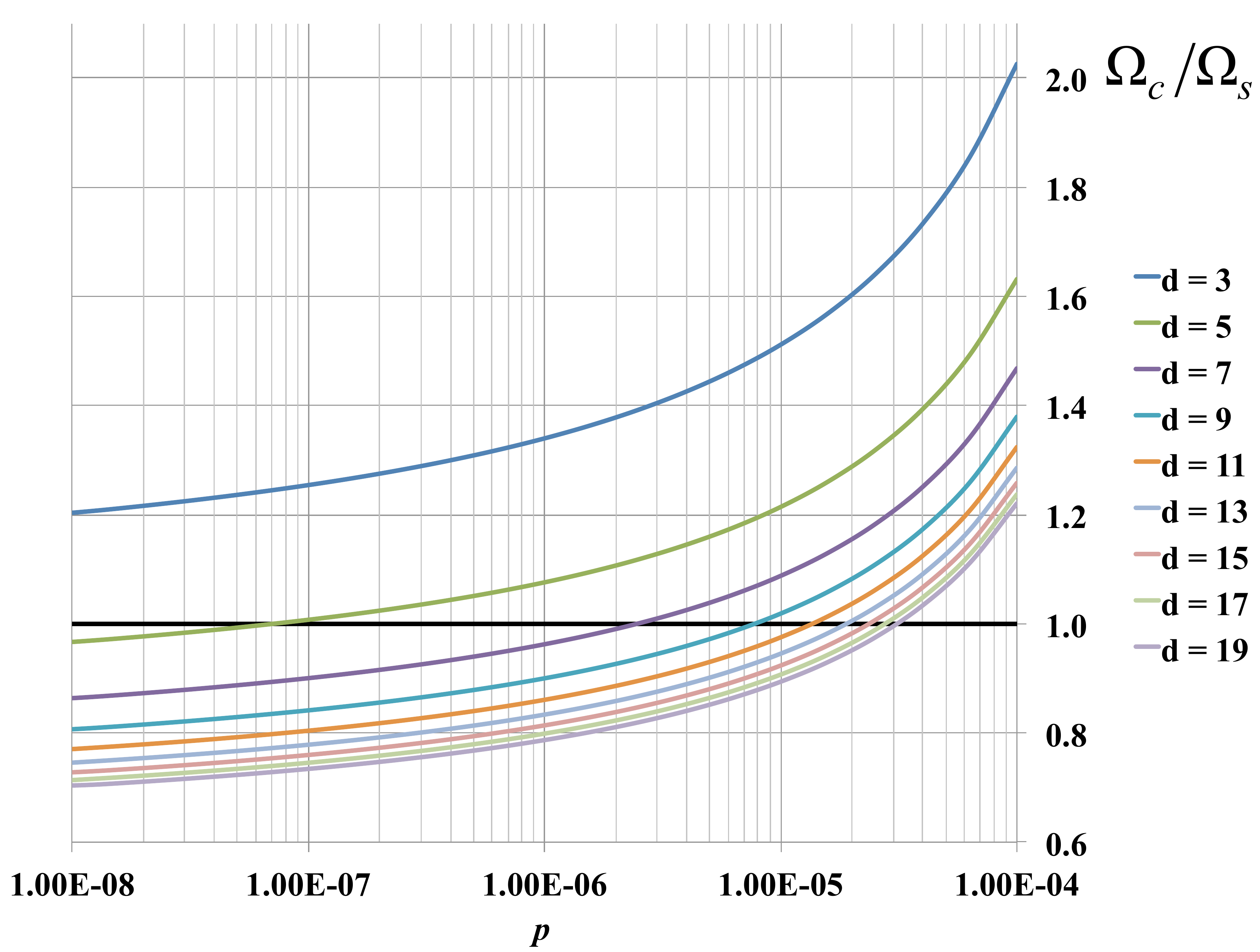}
\caption{\small{\label{fig:overhead-ratio-best-for-color}Ratio of color-code
to surface-code qubit overhead $\Omega_c/\Omega_s$ versus circuit-level
depolarizing probability $p$ when both codes are tuned via
Eq.~(\ref{eq:dc-to-ds}) to achieve the same logical qubit failure
probability.  Plots assume a color-code accuracy threshold of $0.143\%$ and
a surface-code accuracy threshold of $0.502\%$.  (color online)}}
\end{center}
\end{figure}

\begin{figure}[h]
\begin{center}
  \includegraphics[width=1.0\columnwidth]{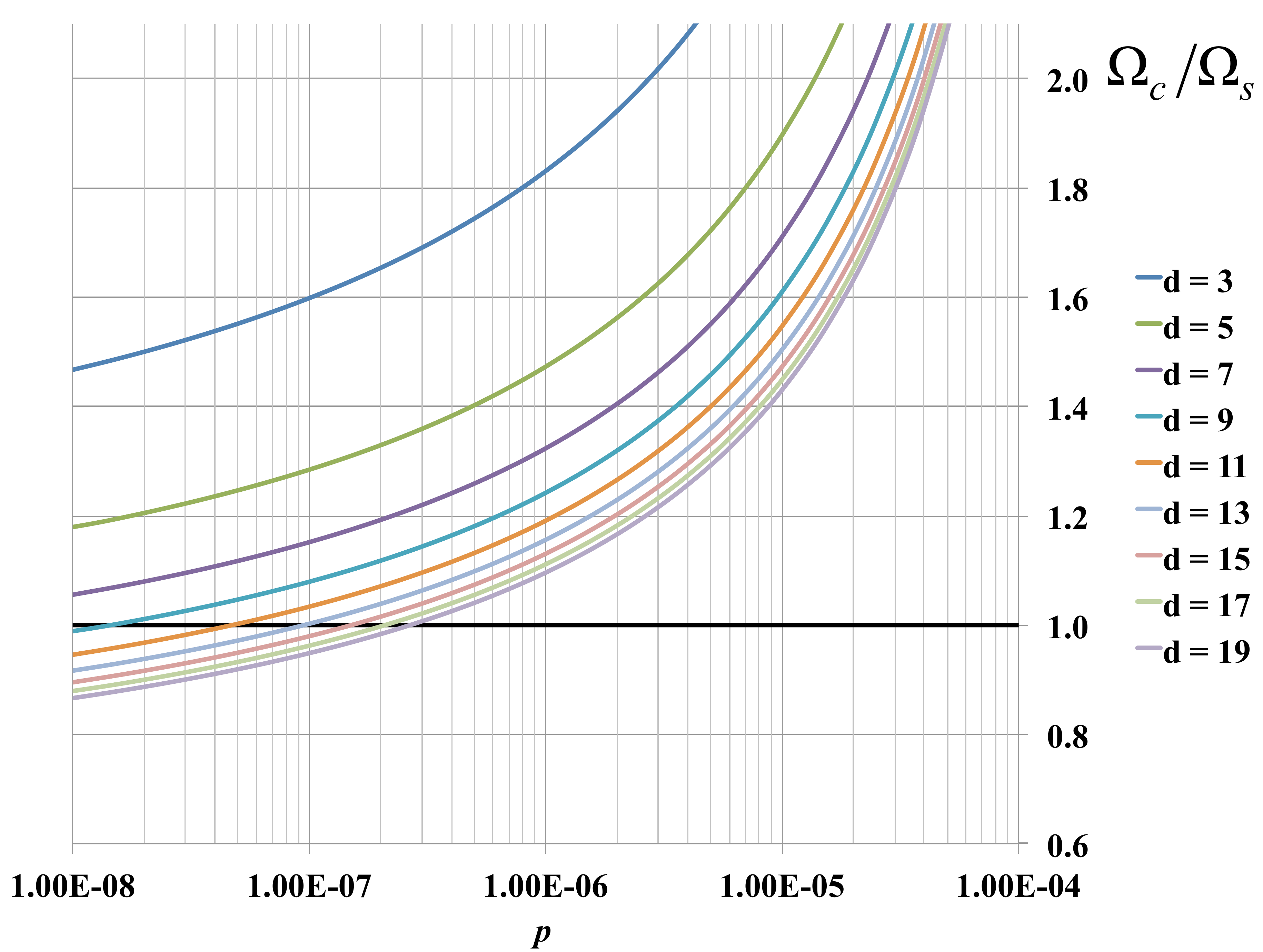}
\caption{\small{\label{fig:overhead-ratio-best-for-surface}Ratio of color-code
to surface-code qubit overhead $\Omega_c/\Omega_s$ versus circuit-level
depolarizing probability $p$ when both codes are tuned via
Eq.~(\ref{eq:dc-to-ds}) to achieve the same logical qubit failure
probability.  Plots assume a color-code accuracy threshold of $0.082\%$ and
a surface-code accuracy threshold of $1.140\%$.  (color online)}}
\end{center}
\end{figure}

Using the expressions in Tables~\ref{tab:color-code-resources-2qubit} and
\ref{tab:surface-code-resources} for the color-code and surface-code qubit
overheads, which we denote by $\Omega_c(d)$ and $\Omega_s(d)$, and the
relationship in Eq.~(\ref{eq:dc-to-ds}), we plot the ratio
$\Omega_c(d_c(d_s))/\Omega_s(d_s)$ versus $p$ for several values of $d_s$.
This ratio is sensitive to the estimates for $p_{\text{th}}^{(c)}$ and
$p_{\text{th}}^{(s)}$, so we present two plots at the extremes of the
estimates. Figure~\ref{fig:overhead-ratio-best-for-color} is the plot using
the highest estimate for the color-code accuracy threshold ($0.143\%$) and
the lowest estimate for the color-code accuracy threshold ($0.502\%$).
Figure~\ref{fig:overhead-ratio-best-for-surface} is the plot using the
lowest estimate for the color-code accuracy threshold ($0.082\%$) and the
highest estimate for the surface-code accuracy threshold ($1.140\%$).

From these plots, we see that for distances greater than 11, as long as $p$
is below a value bracketed approximately somewhere between $10^{-5}$ to
$10^{-7}$, color codes use fewer qubits to achieve the same level of error
suppression.

This conclusion could be sharpened by direct numerical simulations, which we
believe would be an interesting future research project.  Rather than
assuming a phenomenological scaling law as in
Eq.~(\ref{eq:combinatoric-bound}) for the failure probability and using it
to infer the overhead, one could perform direct numerical estimation of the
overhead as a function of $d$ and $p$ and compare the results for color
codes and surface codes.  In addition to removing the need to fit an assumed
scaling law, this approach would also remove the need to estimate accuracy
thresholds, or even pseudothresholds, because it gets directly at the
question at hand.

%
\subsection{Overhead for small logical CNOT gates}
\label{sec:small-CNOTs}

Because of the interest expressed in Ref.~\cite{Horsman:2012a} in designing
the fewest-qubit implementation of a $\CNOT$ gate with a topological
stabilizer code, we thought it would be valuable to list the qubit overheads
required by the methods described here for small distances.  As mentioned in
Sec.~\ref{sec:background}, the number of syndrome qubits used by an
implementation of a topological stabilizer code is design dependent: a
single roving syndrome qubit would suffice, but one could use a number of
syndrome qubits up to twice the number of data qubits to some advantage.

In Table~\ref{tab:CNOT-qubits-small-d}, we list the qubit overhead required
for the low end of the syndrome-allocation spectrum for the following
methods: (a) color-code transversal methods, (b) our color-code
lattice-surgery methods, (c) surface-code transversal methods, (d) our
surface-code lattice-surgery methods, and (e) the surface-code
lattice-surgery methods described in Ref.~\cite{Horsman:2012a}.

As noted in our introduction, transversal methods are not well-suited to
local quantum processing on two-dimensional arrays of qubits restricted to
local movements; we list the overheads here despite this because at small
distances, one might be able to exploit nonlocal processing and/or nonlocal
qubit movement.  For example, a recent demonstration in a trapped-ion
quantum computer of a single-round of error correction on a distance-three
color code exploited the fact that all seven $^{40}\text{Ca}^+$ ions
involved were trapped in a single Paul trap~\cite{Nigg:2014a}.  (The minimal
extra ``roving'' syndrome qubit was not used in the experiment because the
protocol was not fault-tolerant---instead of repeating syndrome measurements
into one or more auxillary qubits, the data-qubit ions were measured
destructively once via resonance-fluorescence.)

For all methods in Table~\ref{tab:CNOT-qubits-small-d}, we consider the
allocations of (i) a single roving syndrome qubit, (ii) one syndrome qubit
per face, and (iii) one syndrome qubit per check, which is the same as one
per face for surface codes but is two per face for color codes.  For
transversal methods, we also consider an in-between variant with (iv) one
syndrome qubit per two faces, because one might want to share the syndrome
qubits transversally between the two logical qubits.  (The case of sharing
two syndrome qubits per face between the two logical qubits has the same
overhead count as having both logical qubits use one syndrome qubit per
face.)


%
\begin{table}[ht!]
\centering
%



\noindent\begin{tabular*}{\columnwidth}{@{\extracolsep{\stretch{1}}}*{6}{l|rrrrr}@{}}
%
\hline\hline
%
%
\hspace{5em} $d$\hspace{1em} & 3 & 5 & 7 & 9 & 11  \\
\hline
Color transversal: 1 total & 15 & 35 & 63 & 99 & 143 \\
%
%
%
Color transversal: $\text{faces}/2$ & 17 & 42 & 77 & 122 & 177 \\
Color transversal: faces & 20 & 50 & 92 & 146 & 212 \\
%
Color transversal: $2\times$faces & 26 & 66 & 122 & 194 & 282 \\
%
%
\hline
Color surgery: 1 total & 22 & 52 & 94 & 148 & 214 \\
Color surgery: faces & 30 & 75 & 138 & 219 & 318 \\
Color surgery: $2\times$faces & 39 & 99 & 183 & 291 & 423 \\
\hline
Surface transversal: 1 total & 19 & 51 & 99 & 163 & 243 \\
Surface transversal: faces/2 & 22 & 66 & 134 & 226 & 342 \\
Surface transversal: faces & 26 & 82 & 170 & 290 & 442 \\
\hline
Surface surgery: 1 total & 28 & 76 & 148 & 244 & 364 \\
Surface surgery: faces & 39 & 123 & 255 & 435 & 663 \\
\hline
Surface surgery~\cite{Horsman:2012a}: 1 total & 34 & 86 & 162 & 262 & 386 \\
Surface surgery~\cite{Horsman:2012a}: faces & 53 & 149 & 293 & 485 & 725 \\
\hline \hline
%
\end{tabular*}
%
\caption{Number of qubits needed to implement a logical $\CNOT$ gate for
several color-code and surface-code methods for small values
of the code distance $d$, assuming that the number of syndrome qubits used
is as indicated.}
\label{tab:CNOT-qubits-small-d}
\end{table}
%

%
\section{Conclusions}
\label{sec:conclusion}

Our color-code lattice-surgery methods open new possibilities for achieving
fault-tolerant quantum computation using fewer resources.  Per code
distance, they are manifestly superior to surface-code lattice-surgery
methods, using approximately half the qubits and the same time or less to
perform logical quantum operations.  Although we did not discuss it, they
also use fewer qubits and the same time or less than defect-based
``spacetime braiding'' methods for both
surface-codes~\cite{Raussendorf:2006a} and color-codes~\cite{Fowler:2008c}.
Transversal methods do use fewer qubits per code distance than color-code
lattice surgery to perform logical operations~\cite{Dennis:2002a}, but
transversal methods cannot be implemented in systems utilizing local quantum
processing on two-dimensional arrays of qubits restricted to local
movements.

Because color codes are estimated to have a lower accuracy threshold than
surface codes against uncorrelated circuit-level depolarizing noise
\cite{Landahl:2011a, Stephens:2014a, Stephens:2014b}, the superiority of
color codes only becomes manifest at sufficiently low depolarizing error
probabilities and sufficiently large code distances.  Subject to an assumed
scaling law given by Eq.~(\ref{eq:combinatoric-bound}) for both surface
codes and color codes, the depolarizing probability cutoff is approximately
somewhere in the range $p = 10^{-5}$ to $p = 10^{-7}$ with a corresponding
distance cutoff of $d = 11$.  Color-code decoder research is only in its
infancy, and we believe that the regime of superiority can be expanded with
further study.  For example, the recent linear-time PEPS decoder by Bravyi
{\textit{et al.}}~\cite{Bravyi:2014a} might be extended to color codes,
allowing one to approximate the optimal decoder quite well with only
linear-time processing.  The close relationship between color codes and
surface codes at the topological-phase level~\cite{Bombin:2012a} means that
the decoding complexity, if not the performance, can always be made
comparable for the two classes of codes~\cite{Bombin:2012a,
Duclos-Cianci:2014a, Delfosse:2014a}.

It would seem then, color codes are equal to or superior to surface codes,
at least insofar as space and time overhead considerations are concerned,
for systems that are sufficiently mature, meaning that they have
sufficiently low error rates and sufficiently many qubits available.  When
technology brings us to this point, we believe the transition from (two-colorable)
surface-codes to (three-colorable) color codes will resemble the transition
of television broadcasts from black-and-white to color: perhaps a little
bumpy at first, but inevitable.  Until then, the mandate for color-code
research is to bring that horizon closer to the present.

%
\begin{acknowledgments}

The authors were supported in part by the Laboratory Directed Research and
Development program at Sandia National Laboratories.  Sandia National
Laboratories is a multi-program laboratory managed and operated by Sandia
Corporation, a wholly owned subsidiary of Lockheed Martin Corporation, for
the U.S.  Department of Energy's  National Nuclear Security Administration
under contract DE-AC04-94AL85000.  The authors would like to thank Eric
Bahr, Chris Cesare, Anand Ganti, Setso Metodi, and Uzoma Onunkwo for helpful
discussions. 

\end{acknowledgments}

%


\bibliographystyle{landahl}
\bibliography{landahl.JabRef}

\begin{thebibliography}{50}
\expandafter\ifx\csname natexlab\endcsname\relax\def\natexlab#1{#1}\fi
\expandafter\ifx\csname bibnamefont\endcsname\relax
  \def\bibnamefont#1{#1}\fi
\expandafter\ifx\csname bibfnamefont\endcsname\relax
  \def\bibfnamefont#1{#1}\fi
\expandafter\ifx\csname citenamefont\endcsname\relax
  \def\citenamefont#1{#1}\fi
\expandafter\ifx\csname url\endcsname\relax
  \def\url#1{\texttt{#1}}\fi
\expandafter\ifx\csname urlprefix\endcsname\relax\def\urlprefix{URL }\fi
\providecommand{\bibinfo}[2]{#2}
\providecommand{\arxiv}[2][]{\href{http://arxiv.org/pdf/#2}{\texttt{arXiv:#2}}}
\providecommand{\doi}[2][]{\href{http://dx.doi.org/#2}{\texttt{doi:#2}}}

\bibitem[{\citenamefont{Stephens}(2014{\natexlab{a}})}]{Stephens:2014b}
\bibinfo{author}{\bibfnamefont{A.~M.} \bibnamefont{Stephens}},
  \emph{\bibinfo{title}{Fault-tolerant thresholds for quantum error correction
  with the surface code}}, \bibinfo{journal}{Phys. Rev. A}
  \textbf{\bibinfo{volume}{89}}, \bibinfo{pages}{022321}
  (\bibinfo{year}{2014}{\natexlab{a}}), \doi{10.1103/PhysRevA.89.022321},
  \arxiv{1311.5003}.

\bibitem[{\citenamefont{Dennis et~al.}(2002)\citenamefont{Dennis, Kitaev,
  Landahl, and Preskill}}]{Dennis:2002a}
\bibinfo{author}{\bibfnamefont{E.}~\bibnamefont{Dennis}},
  \bibinfo{author}{\bibfnamefont{A.}~\bibnamefont{Kitaev}},
  \bibinfo{author}{\bibfnamefont{A.}~\bibnamefont{Landahl}}, \bibnamefont{and}
  \bibinfo{author}{\bibfnamefont{J.}~\bibnamefont{Preskill}},
  \emph{\bibinfo{title}{Topological quantum memory}}, \bibinfo{journal}{J.
  Math. Phys.} \textbf{\bibinfo{volume}{43}}, \bibinfo{pages}{4452}
  (\bibinfo{year}{2002}), \doi{10.1063/1.1499754}, \arxiv{quant-ph/0110143}.

\bibitem[{\citenamefont{Raussendorf and Harrington}(2007)}]{Raussendorf:2007a}
\bibinfo{author}{\bibfnamefont{R.}~\bibnamefont{Raussendorf}} \bibnamefont{and}
  \bibinfo{author}{\bibfnamefont{J.}~\bibnamefont{Harrington}},
  \emph{\bibinfo{title}{Fault-tolerant quantum computation with high threshold
  in two dimensions}}, \bibinfo{journal}{Phys. Rev. Lett.}
  \textbf{\bibinfo{volume}{98}}, \bibinfo{pages}{190504}
  (\bibinfo{year}{2007}), \doi{10.1103/PhysRevLett.98.190504},
  \arxiv{quant-ph/0610082}.

\bibitem[{\citenamefont{Duclos-Cianci and Poulin}(2010)}]{Duclos-Cianci:2009a}
\bibinfo{author}{\bibfnamefont{G.}~\bibnamefont{Duclos-Cianci}}
  \bibnamefont{and} \bibinfo{author}{\bibfnamefont{D.}~\bibnamefont{Poulin}},
  \emph{\bibinfo{title}{Fast decoders for topological quantum codes}},
  \bibinfo{journal}{Phys. Rev. Lett.} \textbf{\bibinfo{volume}{104}},
  \bibinfo{pages}{050504} (\bibinfo{year}{2010}),
  \doi{10.1103/PhysRevLett.104.050504}, \arxiv{0911.0581}.

\bibitem[{\citenamefont{Stephens}(2014{\natexlab{b}})}]{Stephens:2014a}
\bibinfo{author}{\bibfnamefont{A.~M.} \bibnamefont{Stephens}},
  \emph{\bibinfo{title}{Efficient fault-tolerant decoding of topological color
  codes}} (\bibinfo{year}{2014}{\natexlab{b}}), \arxiv{1402.3037}.

\bibitem[{\citenamefont{Cross et~al.}(2009)\citenamefont{Cross, DiVincenzo, and
  Terhal}}]{Cross:2009a}
\bibinfo{author}{\bibfnamefont{A.~W.} \bibnamefont{Cross}},
  \bibinfo{author}{\bibfnamefont{D.~P.} \bibnamefont{DiVincenzo}},
  \bibnamefont{and} \bibinfo{author}{\bibfnamefont{B.~M.}
  \bibnamefont{Terhal}}, \emph{\bibinfo{title}{Comparative code study for
  quantum fault tolerance}}, \bibinfo{journal}{Quant. Inf. Comp.}
  \textbf{\bibinfo{volume}{9}}, \bibinfo{pages}{541} (\bibinfo{year}{2009}),
  \arxiv{0711.1556},
  \urlprefix\url{http://www.rintonpress.com/xxqic9/qic-9-78/0541-0572.pdf}.

\bibitem[{\citenamefont{Anderson}(2011)}]{Anderson:2011a}
\bibinfo{author}{\bibfnamefont{J.~T.} \bibnamefont{Anderson}},
  \emph{\bibinfo{title}{Homological stabilizer codes}} (\bibinfo{year}{2011}),
  \arxiv{1107.2502}.

\bibitem[{\citenamefont{Kitaev}(2003)}]{Kitaev:1997a}
\bibinfo{author}{\bibfnamefont{A.~Y.} \bibnamefont{Kitaev}},
  \emph{\bibinfo{title}{Fault-tolerant quantum computation by anyons}},
  \bibinfo{journal}{Ann. Phys.} \textbf{\bibinfo{volume}{303}},
  \bibinfo{pages}{2} (\bibinfo{year}{2003}),
  \doi{10.1016/S0003-4916(02)00018-0}, \arxiv{quant-ph/9707021}.

\bibitem[{\citenamefont{Bombin and Martin-Delgado}(2006)}]{Bombin:2006b}
\bibinfo{author}{\bibfnamefont{H.}~\bibnamefont{Bombin}} \bibnamefont{and}
  \bibinfo{author}{\bibfnamefont{M.~A.} \bibnamefont{Martin-Delgado}},
  \emph{\bibinfo{title}{Topological quantum distillation}},
  \bibinfo{journal}{Phys. Rev. Lett.} \textbf{\bibinfo{volume}{97}},
  \bibinfo{pages}{180501} (\bibinfo{year}{2006}),
  \doi{10.1103/PhysRevLett.97.180501}, \arxiv{quant-ph/0605138}.

\bibitem[{\citenamefont{Raussendorf et~al.}(2006)\citenamefont{Raussendorf,
  Harrington, and Goyal}}]{Raussendorf:2006a}
\bibinfo{author}{\bibfnamefont{R.}~\bibnamefont{Raussendorf}},
  \bibinfo{author}{\bibfnamefont{J.}~\bibnamefont{Harrington}},
  \bibnamefont{and} \bibinfo{author}{\bibfnamefont{K.}~\bibnamefont{Goyal}},
  \emph{\bibinfo{title}{A fault-tolerant one-way quantum computer}},
  \bibinfo{journal}{Ann. Phys.} \textbf{\bibinfo{volume}{321}},
  \bibinfo{pages}{2242} (\bibinfo{year}{2006}),
  \doi{10.1016/j.aop.2006.01.012}, \arxiv{quant-ph/0510135}.

\bibitem[{\citenamefont{Raussendorf et~al.}(2007)\citenamefont{Raussendorf,
  Harrington, and Goyal}}]{Raussendorf:2007b}
\bibinfo{author}{\bibfnamefont{R.}~\bibnamefont{Raussendorf}},
  \bibinfo{author}{\bibfnamefont{J.}~\bibnamefont{Harrington}},
  \bibnamefont{and} \bibinfo{author}{\bibfnamefont{K.}~\bibnamefont{Goyal}},
  \emph{\bibinfo{title}{Topological fault-tolerance in cluster state quantum
  computation}}, \bibinfo{journal}{New J. Phys.} \textbf{\bibinfo{volume}{9}},
  \bibinfo{pages}{199} (\bibinfo{year}{2007}), \doi{10.1088/1367-2630/9/6/199},
  \arxiv{quant-ph/0703143}.

\bibitem[{\citenamefont{Horsman et~al.}(2012)\citenamefont{Horsman, Fowler,
  Devitt, and {van Meter}}}]{Horsman:2012a}
\bibinfo{author}{\bibfnamefont{C.}~\bibnamefont{Horsman}},
  \bibinfo{author}{\bibfnamefont{A.~G.} \bibnamefont{Fowler}},
  \bibinfo{author}{\bibfnamefont{S.}~\bibnamefont{Devitt}}, \bibnamefont{and}
  \bibinfo{author}{\bibfnamefont{R.}~\bibnamefont{{van Meter}}},
  \emph{\bibinfo{title}{Surface code quantum computing by lattice surgery}},
  \bibinfo{journal}{New J. Phys.} \textbf{\bibinfo{volume}{14}},
  \bibinfo{pages}{123011} (\bibinfo{year}{2012}),
  \doi{10.1088/1367-2630/14/12/123011}, \arxiv{1111.4022}.

\bibitem[{\citenamefont{Spedalieri and Roychowdhury}(2009)}]{Spedalieri:2009a}
\bibinfo{author}{\bibfnamefont{F.~M.} \bibnamefont{Spedalieri}}
  \bibnamefont{and} \bibinfo{author}{\bibfnamefont{V.~P.}
  \bibnamefont{Roychowdhury}}, \emph{\bibinfo{title}{Latency in local,
  two-dimensional, fault-tolerant quantum computing}}, \bibinfo{journal}{Quant.
  Inf. Comp.} \textbf{\bibinfo{volume}{9}}, \bibinfo{pages}{666}
  (\bibinfo{year}{2009}), \arxiv{0805.4213},
  \urlprefix\url{http://www.rintonpress.com/xxqic9/qic-9-78/0666-0682.pdf}.

\bibitem[{\citenamefont{Fowler}(2011)}]{Fowler:2008c}
\bibinfo{author}{\bibfnamefont{A.~G.} \bibnamefont{Fowler}},
  \emph{\bibinfo{title}{Two-dimensional color-code quantum computation}},
  \bibinfo{journal}{Phys. Rev. A} \textbf{\bibinfo{volume}{83}},
  \bibinfo{pages}{042310} (\bibinfo{year}{2011}),
  \doi{10.1103/PhysRevA.83.042310}, \arxiv{0806.4827v3}.

\bibitem[{\citenamefont{Bombin and Martin-Delgado}(2007)}]{Bombin:2007d}
\bibinfo{author}{\bibfnamefont{H.}~\bibnamefont{Bombin}} \bibnamefont{and}
  \bibinfo{author}{\bibfnamefont{M.~A.} \bibnamefont{Martin-Delgado}},
  \emph{\bibinfo{title}{Optimal resources for topological two-dimensional
  stabilizer codes: Comparative study}}, \bibinfo{journal}{Phys. Rev. A}
  \textbf{\bibinfo{volume}{76}}, \bibinfo{pages}{012305}
  (\bibinfo{year}{2007}), \doi{10.1103/PhysRevA.76.012305},
  \arxiv{quant-ph/0703272}.

\bibitem[{\citenamefont{Landahl et~al.}(2011)\citenamefont{Landahl, Anderson,
  and Rice}}]{Landahl:2011a}
\bibinfo{author}{\bibfnamefont{A.~J.} \bibnamefont{Landahl}},
  \bibinfo{author}{\bibfnamefont{J.~T.} \bibnamefont{Anderson}},
  \bibnamefont{and} \bibinfo{author}{\bibfnamefont{P.~R.} \bibnamefont{Rice}},
  \emph{\bibinfo{title}{Fault-tolerant quantum computing with color codes}}
  (\bibinfo{year}{2011}), \arxiv{1108.5738}.

\bibitem[{\citenamefont{Fowler}(2012)}]{Fowler:2012c}
\bibinfo{author}{\bibfnamefont{A.~G.} \bibnamefont{Fowler}},
  \emph{\bibinfo{title}{Low-overhead surface code logical {H}adamard}},
  \bibinfo{journal}{Quant. Inf. Comp.} \textbf{\bibinfo{volume}{12}},
  \bibinfo{pages}{970} (\bibinfo{year}{2012}), \arxiv{1202.2639},
  \urlprefix\url{http://www.rintonpress.com/xxqic12/qic-12-1112/0970-0982.pdf}.

\bibitem[{\citenamefont{Aliferis}(2007)}]{Aliferis:2007b}
\bibinfo{author}{\bibfnamefont{P.}~\bibnamefont{Aliferis}},
  \emph{\bibinfo{title}{Level reduction and the quantum threshold theorem}},
  Ph.D. thesis, \bibinfo{school}{Caltech} (\bibinfo{year}{2007}),
  \arxiv{quant-ph/0703230}.

\bibitem[{\citenamefont{Gottesman}(1997)}]{Gottesman:1997a}
\bibinfo{author}{\bibfnamefont{D.}~\bibnamefont{Gottesman}},
  \emph{\bibinfo{title}{Stabilizer codes and quantum error correction}}, Ph.D.
  thesis, \bibinfo{school}{Caltech} (\bibinfo{year}{1997}),
  \arxiv{quant-ph/9705052}.

\bibitem[{\citenamefont{Shor}(1996)}]{Shor:1996a}
\bibinfo{author}{\bibfnamefont{P.~W.} \bibnamefont{Shor}},
  \emph{\bibinfo{title}{Fault-tolerant quantum computation}}, in
  \emph{\bibinfo{booktitle}{Proceedings of the 37th Annual Symposium on
  Foundations of Computer Science}}, edited by
  \bibinfo{editor}{\bibfnamefont{R.~S.} \bibnamefont{Sipple}},
  \bibinfo{organization}{IEEE} (\bibinfo{publisher}{IEEE Press, Los Alamitos,
  CA}, \bibinfo{address}{14--16 Oct. 1996, Burlington, VT, USA},
  \bibinfo{year}{1996}), pp. \bibinfo{pages}{56--65}, ISBN
  \bibinfo{isbn}{0-8186-7594-2}, \doi{10.1137/S0097539795293172},
  \arxiv{quant-ph/9605011}.

\bibitem[{\citenamefont{Steane}(1998)}]{Steane:1998a}
\bibinfo{author}{\bibfnamefont{A.~M.} \bibnamefont{Steane}},
  \emph{\bibinfo{title}{Space, time, parallelism and noise requirements for
  reliable quantum computing}}, \bibinfo{journal}{Fortschr. Phys.}
  \textbf{\bibinfo{volume}{46}}, \bibinfo{pages}{443} (\bibinfo{year}{1998}),
  \arxiv{quant-ph/9708021},
  \urlprefix\url{http://onlinelibrary.wiley.com/doi/10.1002/%28SICI%291521-3978%28199806%2946:4/5%3C443::AID-PROP443%3E3.0.CO;2-8/abstract}.

\bibitem[{\citenamefont{Knill}(2005)}]{Knill:2004a}
\bibinfo{author}{\bibfnamefont{E.}~\bibnamefont{Knill}},
  \emph{\bibinfo{title}{Quantum computing with realistically noisy devices}},
  \bibinfo{journal}{Nature} \textbf{\bibinfo{volume}{434}}, \bibinfo{pages}{39}
  (\bibinfo{year}{2005}), \doi{10.1038/nature03350}, \arxiv{quant-ph/0410199}.

\bibitem[{\citenamefont{Magesan et~al.}(2013)\citenamefont{Magesan, Puzzuoli,
  Granade, and Cory}}]{Magesan:2013a}
\bibinfo{author}{\bibfnamefont{E.}~\bibnamefont{Magesan}},
  \bibinfo{author}{\bibfnamefont{D.}~\bibnamefont{Puzzuoli}},
  \bibinfo{author}{\bibfnamefont{C.~E.} \bibnamefont{Granade}},
  \bibnamefont{and} \bibinfo{author}{\bibfnamefont{D.~G.} \bibnamefont{Cory}},
  \emph{\bibinfo{title}{Modeling quantum noise for efficient testing of
  fault-tolerant circuits}}, \bibinfo{journal}{Phys. Rev. A}
  \textbf{\bibinfo{volume}{87}}, \bibinfo{pages}{012324}
  (\bibinfo{year}{2013}), \doi{10.1103/PhysRevA.87.012324}, \arxiv{1206.5407}.

\bibitem[{\citenamefont{Gutierrez et~al.}(2013)\citenamefont{Gutierrez, Svec,
  Vargo, and Brown}}]{Gutierrez:2013a}
\bibinfo{author}{\bibfnamefont{M.}~\bibnamefont{Gutierrez}},
  \bibinfo{author}{\bibfnamefont{L.}~\bibnamefont{Svec}},
  \bibinfo{author}{\bibfnamefont{A.}~\bibnamefont{Vargo}}, \bibnamefont{and}
  \bibinfo{author}{\bibfnamefont{K.~R.} \bibnamefont{Brown}},
  \emph{\bibinfo{title}{Approximation of real error channels by {C}lifford
  channels and {P}auli measurements}}, \bibinfo{journal}{Phys. Rev. A}
  \textbf{\bibinfo{volume}{87}}, \bibinfo{pages}{030302}
  (\bibinfo{year}{2013}), \doi{10.1103/PhysRevA.87.030302}, \arxiv{1207.0046}.

\bibitem[{\citenamefont{Iyer and Poulin}(2013)}]{Iyer:2013a}
\bibinfo{author}{\bibfnamefont{P.}~\bibnamefont{Iyer}} \bibnamefont{and}
  \bibinfo{author}{\bibfnamefont{D.}~\bibnamefont{Poulin}},
  \emph{\bibinfo{title}{Hardness of decoding quantum stabilizer codes}}
  (\bibinfo{year}{2013}), \arxiv{1310.3235}.

\bibitem[{\citenamefont{Bravyi et~al.}(2014)\citenamefont{Bravyi, Suchara, and
  Vargo}}]{Bravyi:2014a}
\bibinfo{author}{\bibfnamefont{S.}~\bibnamefont{Bravyi}},
  \bibinfo{author}{\bibfnamefont{M.}~\bibnamefont{Suchara}}, \bibnamefont{and}
  \bibinfo{author}{\bibfnamefont{A.}~\bibnamefont{Vargo}},
  \emph{\bibinfo{title}{Efficient algorithms for maximum likelihood decoding in
  the surface code}} (\bibinfo{year}{2014}), \arxiv{1405.4883}.

\bibitem[{\citenamefont{Wang et~al.}(2010)\citenamefont{Wang, Fowler, Hill, and
  Hollenberg}}]{Wang:2009b}
\bibinfo{author}{\bibfnamefont{D.~S.} \bibnamefont{Wang}},
  \bibinfo{author}{\bibfnamefont{A.~G.} \bibnamefont{Fowler}},
  \bibinfo{author}{\bibfnamefont{C.~D.} \bibnamefont{Hill}}, \bibnamefont{and}
  \bibinfo{author}{\bibfnamefont{L.~C.~L.} \bibnamefont{Hollenberg}},
  \emph{\bibinfo{title}{Graphical algorithms and threshold error rates for the
  2d color code}}, \bibinfo{journal}{Quant. Inf. Comp.}
  \textbf{\bibinfo{volume}{10}}, \bibinfo{pages}{780} (\bibinfo{year}{2010}),
  \arxiv{0907.1708},
  \urlprefix\url{http://www.rintonpress.com/xxqic10/qic-10-910/0780-0802.pdf}.

\bibitem[{\citenamefont{Sarvepalli and Raussendorf}(2011)}]{Sarvepalli:2011a}
\bibinfo{author}{\bibfnamefont{P.}~\bibnamefont{Sarvepalli}} \bibnamefont{and}
  \bibinfo{author}{\bibfnamefont{R.}~\bibnamefont{Raussendorf}},
  \emph{\bibinfo{title}{Efficient decoding of topological color codes}}
  (\bibinfo{year}{2011}), \arxiv{1111.0831}.

\bibitem[{\citenamefont{Duclos-Cianci et~al.}(2011)\citenamefont{Duclos-Cianci,
  Bomb{\'{i}}n, and Poulin}}]{Duclos-Cianci:2011a}
\bibinfo{author}{\bibfnamefont{G.}~\bibnamefont{Duclos-Cianci}},
  \bibinfo{author}{\bibfnamefont{H.}~\bibnamefont{Bomb{\'{i}}n}},
  \bibnamefont{and} \bibinfo{author}{\bibfnamefont{D.}~\bibnamefont{Poulin}},
  \emph{\bibinfo{title}{Fast decoding algorithm for subspace and subsystem
  color codes and local equivalence of topological phases}}
  (\bibinfo{year}{2011}), \bibinfo{note}{personal communication.}

\bibitem[{\citenamefont{Duclos-Cianci and Poulin}(2014)}]{Duclos-Cianci:2014a}
\bibinfo{author}{\bibfnamefont{G.}~\bibnamefont{Duclos-Cianci}}
  \bibnamefont{and} \bibinfo{author}{\bibfnamefont{D.}~\bibnamefont{Poulin}},
  \emph{\bibinfo{title}{Fault-tolerant renormalization group decoder for
  {A}belian topological codes}}, \bibinfo{journal}{Quant. Inf. Comp.}
  \textbf{\bibinfo{volume}{14}}, \bibinfo{pages}{0721} (\bibinfo{year}{2014}),
  \arxiv{1304.6100},
  \urlprefix\url{http://www.rintonpress.com/xxqic14/qic-14-910/0721-0740.pdf}.

\bibitem[{\citenamefont{Dennis}(2003)}]{Dennis:2003a}
\bibinfo{author}{\bibfnamefont{E.}~\bibnamefont{Dennis}},
  \emph{\bibinfo{title}{Purifying quantum states: quantum and classical
  algorithms}}, Ph.D. thesis, \bibinfo{school}{University of California at
  Santa Barbara} (\bibinfo{year}{2003}), \arxiv{quant-ph/0503169}.

\bibitem[{\citenamefont{Bravyi and Vargo}(2013)}]{Bravyi:2013a}
\bibinfo{author}{\bibfnamefont{S.}~\bibnamefont{Bravyi}} \bibnamefont{and}
  \bibinfo{author}{\bibfnamefont{A.}~\bibnamefont{Vargo}},
  \emph{\bibinfo{title}{Simulation of rare events in quantum error
  correction}}, \bibinfo{journal}{Phys. Rev. A} \textbf{\bibinfo{volume}{88}},
  \bibinfo{pages}{062308} (\bibinfo{year}{2013}),
  \doi{10.1103/PhysRevA.88.062308}, \arxiv{1308.6270}.

\bibitem[{\citenamefont{Wootton}(2013)}]{Wootton:2013a}
\bibinfo{author}{\bibfnamefont{J.~R.} \bibnamefont{Wootton}},
  \emph{\bibinfo{title}{A simple decoder for topological codes}}
  (\bibinfo{year}{2013}), \arxiv{1310.2393}.

\bibitem[{\citenamefont{Anwar et~al.}(2014)\citenamefont{Anwar, Brown,
  Campbell, and Browne}}]{Anwar:2014a}
\bibinfo{author}{\bibfnamefont{H.}~\bibnamefont{Anwar}},
  \bibinfo{author}{\bibfnamefont{B.~J.} \bibnamefont{Brown}},
  \bibinfo{author}{\bibfnamefont{E.~T.} \bibnamefont{Campbell}},
  \bibnamefont{and} \bibinfo{author}{\bibfnamefont{D.~E.}
  \bibnamefont{Browne}}, \emph{\bibinfo{title}{Efficient decoders for qudit
  topological codes}}, \bibinfo{journal}{New J. Phys.}
  \textbf{\bibinfo{volume}{16}}, \bibinfo{pages}{063038}
  (\bibinfo{year}{2014}), \doi{10.1088/1367-2630/16/6/063038},
  \arxiv{1311.4895}.

\bibitem[{\citenamefont{Harrington}(2004)}]{Harrington:2004a}
\bibinfo{author}{\bibfnamefont{J.~W.} \bibnamefont{Harrington}},
  \emph{\bibinfo{title}{Analysis of quantum error-correcting codes: symplectic
  lattice codes and toric codes}}, Ph.D. thesis, \bibinfo{school}{Caltech}
  (\bibinfo{year}{2004}).

\bibitem[{\citenamefont{Herold et~al.}(2014)\citenamefont{Herold, Campbell,
  Eisert, and Kastoryano}}]{Herold:2014a}
\bibinfo{author}{\bibfnamefont{M.}~\bibnamefont{Herold}},
  \bibinfo{author}{\bibfnamefont{E.~T.} \bibnamefont{Campbell}},
  \bibinfo{author}{\bibfnamefont{J.}~\bibnamefont{Eisert}}, \bibnamefont{and}
  \bibinfo{author}{\bibfnamefont{M.~J.} \bibnamefont{Kastoryano}},
  \emph{\bibinfo{title}{Cellular-automaton decoders for topological quantum
  memories}} (\bibinfo{year}{2014}), \arxiv{1406.2338}.

\bibitem[{\citenamefont{Bombin et~al.}(2012)\citenamefont{Bombin,
  Duclos-Cianci, and Poulin}}]{Bombin:2012a}
\bibinfo{author}{\bibfnamefont{H.}~\bibnamefont{Bombin}},
  \bibinfo{author}{\bibfnamefont{G.}~\bibnamefont{Duclos-Cianci}},
  \bibnamefont{and} \bibinfo{author}{\bibfnamefont{D.}~\bibnamefont{Poulin}},
  \emph{\bibinfo{title}{Universal topological phase of 2{D} stabilizer codes}},
  \bibinfo{journal}{New J. Phys.} \textbf{\bibinfo{volume}{14}},
  \bibinfo{pages}{073048} (\bibinfo{year}{2012}), \arxiv{1103.4606}.

\bibitem[{\citenamefont{Delfosse}(2014)}]{Delfosse:2014a}
\bibinfo{author}{\bibfnamefont{N.}~\bibnamefont{Delfosse}},
  \emph{\bibinfo{title}{Decoding color codes by projection onto surface
  codes}}, \bibinfo{journal}{Phys. Rev. A} \textbf{\bibinfo{volume}{89}},
  \bibinfo{pages}{012317} (\bibinfo{year}{2014}),
  \doi{10.1103/PhysRevA.89.012317}, \arxiv{1308.6207}.

\bibitem[{\citenamefont{Nielsen and Chuang}(2000)}]{Nielsen:2000a}
\bibinfo{author}{\bibfnamefont{M.~A.} \bibnamefont{Nielsen}} \bibnamefont{and}
  \bibinfo{author}{\bibfnamefont{I.~L.} \bibnamefont{Chuang}},
  \emph{\bibinfo{title}{Quantum Computation and Quantum Information}}
  (\bibinfo{publisher}{Cambridge University Press},
  \bibinfo{address}{Cambridge}, \bibinfo{year}{2000}), ISBN
  \bibinfo{isbn}{0-521-63235-8 (Hardback), 0-521-63503-9 (Paperback)}.

\bibitem[{\citenamefont{Gottesman}(1999)}]{Gottesman:1999b}
\bibinfo{author}{\bibfnamefont{D.}~\bibnamefont{Gottesman}},
  \emph{\bibinfo{title}{The {H}eisenberg representation of quantum computers}},
  in \emph{\bibinfo{booktitle}{Group22: Proceedings of the XXII International
  Colloquium on Group Theoretical Methods in Physics}}, edited by
  \bibinfo{editor}{\bibfnamefont{S.~P.} \bibnamefont{Corney}},
  \bibinfo{editor}{\bibfnamefont{R.}~\bibnamefont{Delbourgo}},
  \bibnamefont{and} \bibinfo{editor}{\bibfnamefont{P.~D.} \bibnamefont{Jarvis}}
  (\bibinfo{publisher}{International Press, Cambridge, MA},
  \bibinfo{address}{13--17 Jul. 1998, Hobart, Australia},
  \bibinfo{year}{1999}), pp. \bibinfo{pages}{32--43}, \arxiv{quant-ph/9807006}.

\bibitem[{\citenamefont{Bravyi and Kitaev}(2005)}]{Bravyi:2005a}
\bibinfo{author}{\bibfnamefont{S.}~\bibnamefont{Bravyi}} \bibnamefont{and}
  \bibinfo{author}{\bibfnamefont{A.}~\bibnamefont{Kitaev}},
  \emph{\bibinfo{title}{Universal quantum computation with ideal {C}lifford
  gates and noisy ancillas}}, \bibinfo{journal}{Phys. Rev. A}
  \textbf{\bibinfo{volume}{71}}, \bibinfo{pages}{022316}
  (\bibinfo{year}{2005}), \doi{10.1103/PhysRevA.71.022316},
  \arxiv{quant-ph/0403025}.

\bibitem[{\citenamefont{Meier et~al.}(2012)\citenamefont{Meier, Eastin, and
  Knill}}]{Meier:2012a}
\bibinfo{author}{\bibfnamefont{A.~M.} \bibnamefont{Meier}},
  \bibinfo{author}{\bibfnamefont{B.}~\bibnamefont{Eastin}}, \bibnamefont{and}
  \bibinfo{author}{\bibfnamefont{E.}~\bibnamefont{Knill}},
  \emph{\bibinfo{title}{Magic-state distillation with the four-qubit code}}
  (\bibinfo{year}{2012}), \arxiv{1204.4221}.

\bibitem[{\citenamefont{Bravyi and Haah}(2012)}]{Bravyi:2012a}
\bibinfo{author}{\bibfnamefont{S.}~\bibnamefont{Bravyi}} \bibnamefont{and}
  \bibinfo{author}{\bibfnamefont{J.}~\bibnamefont{Haah}},
  \emph{\bibinfo{title}{Magic state distillation with low overhead}}
  (\bibinfo{year}{2012}), \arxiv{1209.2426}.

\bibitem[{\citenamefont{Landahl and Cesare}(2013)}]{Landahl:2013a}
\bibinfo{author}{\bibfnamefont{A.~J.} \bibnamefont{Landahl}} \bibnamefont{and}
  \bibinfo{author}{\bibfnamefont{C.}~\bibnamefont{Cesare}},
  \emph{\bibinfo{title}{Complex instruction set computing architecture for
  performing accurate quantum $z$ rotations with less magic}}
  (\bibinfo{year}{2013}), \arxiv{1302.3240}.

\bibitem[{\citenamefont{Bombin}(2013)}]{Bombin:2013a}
\bibinfo{author}{\bibfnamefont{H.}~\bibnamefont{Bombin}},
  \emph{\bibinfo{title}{Gauge color codes}} (\bibinfo{year}{2013}),
  \arxiv{1311.0879}.

\bibitem[{\citenamefont{Aliferis and Preskill}(2008)}]{Aliferis:2008b}
\bibinfo{author}{\bibfnamefont{P.}~\bibnamefont{Aliferis}} \bibnamefont{and}
  \bibinfo{author}{\bibfnamefont{J.}~\bibnamefont{Preskill}},
  \emph{\bibinfo{title}{Fault-tolerant quantum computation against biased
  noise}}, \bibinfo{journal}{Phys. Rev. A} \textbf{\bibinfo{volume}{78}},
  \bibinfo{pages}{052331} (\bibinfo{year}{2008}),
  \doi{10.1103/PhysRevA.78.052331}, \arxiv{0710.1301}.

\bibitem[{\citenamefont{Wang et~al.}(2003)\citenamefont{Wang, Harrington, and
  Preskill}}]{Wang:2003a}
\bibinfo{author}{\bibfnamefont{C.}~\bibnamefont{Wang}},
  \bibinfo{author}{\bibfnamefont{J.}~\bibnamefont{Harrington}},
  \bibnamefont{and} \bibinfo{author}{\bibfnamefont{J.}~\bibnamefont{Preskill}},
  \emph{\bibinfo{title}{Confinement-{H}iggs transition in a disordered gauge
  theory and the accuracy threshold for quantum memory}},
  \bibinfo{journal}{Ann. Phys.} \textbf{\bibinfo{volume}{303}},
  \bibinfo{pages}{31} (\bibinfo{year}{2003}),
  \doi{10.1016/S0003-4916(02)00019-2}, \arxiv{quant-ph/0207088}.

\bibitem[{\citenamefont{Fowler}(2013)}]{Fowler:2013a}
\bibinfo{author}{\bibfnamefont{A.~G.} \bibnamefont{Fowler}},
  \emph{\bibinfo{title}{Analytic asymptotic performance of topological codes}},
  \bibinfo{journal}{Phys. Rev. A} \textbf{\bibinfo{volume}{87}},
  \bibinfo{pages}{040310(R)} (\bibinfo{year}{2013}),
  \doi{10.1103/PhysRevA.87.040301}, \arxiv{1208.1334}.

\bibitem[{\citenamefont{Watson and Barrett}(2013)}]{Watson:2013a}
\bibinfo{author}{\bibfnamefont{F.~H.} \bibnamefont{Watson}} \bibnamefont{and}
  \bibinfo{author}{\bibfnamefont{S.~D.} \bibnamefont{Barrett}},
  \emph{\bibinfo{title}{Estimating overhead for topological quantum codes}}
  (\bibinfo{year}{2013}), \arxiv{1312.5213}.

\bibitem[{\citenamefont{Nigg et~al.}(2014)\citenamefont{Nigg, Mueller,
  Martinez, Schindler, Hennrich, Monz, Martin-Delgado, and Blatt}}]{Nigg:2014a}
\bibinfo{author}{\bibfnamefont{D.}~\bibnamefont{Nigg}},
  \bibinfo{author}{\bibfnamefont{M.}~\bibnamefont{Mueller}},
  \bibinfo{author}{\bibfnamefont{E.~A.} \bibnamefont{Martinez}},
  \bibinfo{author}{\bibfnamefont{P.}~\bibnamefont{Schindler}},
  \bibinfo{author}{\bibfnamefont{M.}~\bibnamefont{Hennrich}},
  \bibinfo{author}{\bibfnamefont{T.}~\bibnamefont{Monz}},
  \bibinfo{author}{\bibfnamefont{M.~A.} \bibnamefont{Martin-Delgado}},
  \bibnamefont{and} \bibinfo{author}{\bibfnamefont{R.}~\bibnamefont{Blatt}},
  \emph{\bibinfo{title}{Experimental quantum computations on a topologcially
  encoded qubit}} (\bibinfo{year}{2014}), \arxiv{1403.5426}.

\end{thebibliography}

\end{document}